\documentclass[aps,prr,reprint,amsmath,floatfix,amssymb,superscriptaddress]{revtex4-2}
\usepackage{graphicx}
\usepackage{dcolumn}
\usepackage{bm}
\usepackage{color}
\usepackage{hyperref}
\hypersetup{colorlinks,allcolors=blue}
\usepackage{lipsum}

\newcommand{\be}{\begin{equation}}
\newcommand{\ee}{\end{equation}}
\newcommand{\bea}{\begin{eqnarray}}
\newcommand{\eea}{\end{eqnarray}}
\newcommand{\ba}[1]{\begin{array}{#1}}
\newcommand{\ea}{\end{array}}

\begin{document}


\title{Efficient excitation transfer in an LH2-inspired nanoscale stacked ring geometry}
\author{Arpita Pal}
\thanks{Contact author: Arpita.Pal@uibk.ac.at}
\affiliation{Institut für Theoretische Physik, Universität Innsbruck, Technikerstraße 21a, A-6020 Innsbruck, Austria}
\author{Raphael Holzinger}
\affiliation{Institut für Theoretische Physik, Universität Innsbruck, Technikerstraße 21a, A-6020 Innsbruck, Austria}
\author{Maria Moreno-Cardoner}
\affiliation{Departament de Física Quàntica i Astrofísica and Institut de Ciències del Cosmos, Universitat de Barcelona, Martí i Franquès 1, E-08028 Barcelona, Spain}
\author{Helmut Ritsch}
\affiliation{Institut für Theoretische Physik, Universität Innsbruck, Technikerstraße 21a, A-6020 Innsbruck, Austria}
\date{\today}

\begin{abstract}
Subwavelength ring-shaped structures of quantum emitters exhibit outstanding radiation properties and are useful for antennas, excitation transport, and storage. Taking inspiration from the oligomeric geometry of biological light-harvesting 2 (LH2) complexes, we study here generic examples and predict highly efficient excitation transfer in a three-dimensional (3D) subwavelength concentric stacked ring structure with a diameter of 400 \textit{nm}, formed by two-level atoms. Utilizing the quantum optical open system master equation approach for the collective dipole dynamics, we demonstrate that, depending on the system parameters, our bio-mimicked 3D ring enables efficient excitation transfer between two ring layers. Our findings open prospects for engineering other biomimetic light-matter platforms and emitter arrays to achieve efficient energy transfer.
\end{abstract}
\maketitle

\section{Introduction}
With the recent technological advances in position and optical control of quantum emitters even at sub-wavelength distances in a variety of platforms~\cite{barredo:science:2016, Hemmig:nl:2016, barredo:nature:2018,Dalacu_2019, rui2020subradiant, lounis:nc:2022, lodhal:science:2023, Hood:np:2024, Thompson:nature:2012, lukin:nature:2020, kaufman:np:2021, Ohmori:np:2022, kirchmair:np:2022}, the interest in their intrinsic collective dynamics~\cite{dicke:1954,haroche:1982, vahid:science:2002, hommel:np:2007, zoubi:pra:2011, gauger:nc:2014,plenio:NJP:2014,kimble:prl:2015,Zelevinsky:np:2015,cirac:prl:2015,kaiser:prl:2016,Ruostekoski:prl:2016,guerin:prl:2016,lukin:prl:2017,asenjo:prx:2017,shahmoon:prl:2017,solano:nc:2017,laurat:nature:2019,Mariona:pra:2019,zoller:prl:2019,hughes:prb:2019,rauschenbeutel:nphoton:2020,yelin:prl:2021,Ferioli:prx:2021,masson:nc:2022,genes:prx:2022, garcia:2022:prr, huidobro:prr:2023, Janne:PRA:2023} and possible applications in quantum information~\cite{hammerer:rmp:2010}, and quantum sensing~\cite{zoller:prl:2024} have triggered a renewed wave of theoretical modeling. In particular, the emerging manifold of sub-radiant states~\cite{brewer:prl:1996, laurin:prl:2013, genes:sr:2015, scully:prl:2015, robicheaux:pra:2016, kaiser:prl:2016, Ruostekoski:prl:2016, molmer:prl:2019, Mariona:pra:2019, Needham_2019, Ferioli:prx:2021, oriol:prr:2022, lounis:nc:2022, lodhal:science:2023, Hood:np:2024} has a high potential to engineer and control coherent dynamics at extremely low loss~\cite{rui2020subradiant} and even implement nonlinear elements~\cite{holzinger2021nanoscale, moreno:prl:2021,  Mariona2022Efficient}. Interestingly sub-wavelength ring structures are also omnipresent in biological light-harvesting (LH) complexes of purple photosynthetic bacteria~\cite{McDermott:nature:1995, KUHLBRANDT:1995, koepke:structure:1996} with their precise functionality despite extensive quantum-chemical modeling not fully understood yet. In a much-simplified quantum optical analog, the presence of finite dipole moments in each bacteriochlorophyll's (BChl's) enables cooperative quantum activity to explain outstanding optical properties of LH complexes~\cite{plenio:NJP:2014, Mariona:pra:2019, Needham_2019, Cremer:NJP:2020, holzinger2021nanoscale, genes:prx:2022, Mariona2022Efficient, Janne:PRA:2023, verena:nm:2023, Holzinger2024harnessing}. Specifically, a model nonameric ring with a central absorber shows superior light absorption properties~\cite{Mariona2022Efficient} even in the presence of phonons~\cite{holzinger2022cooperative} and enables excitation transport~\cite{Mariona:pra:2019} robust to noise over longer ring chains~\cite{Holzinger2024harnessing}. 

Ring-shaped light-harvesting (LH) aggregates, in particular, LH1 and LH2 complexes exhibit circular arrangements of multiple pigments with certain rotational symmetries~\cite{KUHLBRANDT:1995,McDermott:nature:1995,koepke:structure:1996,Kohler:revbio:2006,Yu:nature:2018}. In photosynthesis, the coupled ring geometries of LH complexes essentially result in a highly efficient light capture and extremely fast excitation transfer mechanism~\cite{VANGRONDELLE:1994,schulten:cpc:2011,Mirkovic:cr:2017,book:photosynthetic_excitons,vladimir:pccp:2006,kassal:jcp:2013,mennucci:jpcl:2018,vladimir:pccp:2023,maity:review:2023}. A network of coupled rings facilitates a near loss-less propagation of light energy towards the reaction center~\cite{Mohseni_Omar_Engel_Plenio_2014}. In the LH2 complex nature has designed two circular stacked layers of BChls (with sparse and dense aggregation), which simultaneously exhibit intra-ring and inter-layer~\cite{scholes:jpcb:2000,vanHulst:science:2013,castro:arxiv:2024} excitation transport mechanisms~\cite{FLEMING:structure:1997}. Understanding nature's design principles for light harvesting~\cite{Scholes:nc:2011,zaks:fd:2012,Sarovar_2013} and implementing them in artificial nanoscale devices~\cite{Kundu:cr:2017,gauger:2024:spie} are of crucial importance, as this opens up possibilities for engineering realistic ring-shaped devices that inherit this high-efficiency~\cite{Parkinson:JACS:2014, holzinger:prl:2020, Mariona2022Efficient, gauger:prx:2023,
Rueda:jpcc:2024}. Such efforts also offer a promising route to optimize energy transport~\cite{olmos:pra:2010, Mariona:pra:2019, Cremer:NJP:2020, Han:njp:2023, Amico:pra:2023, Amico:pra:2024, Holzinger2024harnessing} through robust and resilient ring architectures, which could provide unforeseen benefits for sustainable quantum technologies~\cite{Alexia:prxq:2022}.

Here we take a deeper look into the arrangement of BChls in a single LH2 complex to uncover some of nature's secrets in ring design and understand its' influence on efficient excitation energy transfer using the Markovian master equation approach for a single photon manifold. This should allow us to optimize excitation transfer in both biological geometries and larger biomimetic  stacked nanorings. We utilize the geometry of the most abundant nonameric LH2 complex of \textit{Rhodoblastus (Rbl.) acidophilus} (previously recognized as \textit{Rhodopseudomonas acidophila})~\cite{McDermott:nature:1995, Kohler:revbio:2006} [Fig.~\ref{fig1}(a)] for quantum optical analysis. A logical alteration of the layer separation together with considering all BChl to be photoactive at the same wavelength, would theoretically estimate for highly efficient inter-layer excitation energy transfer in the sub-nano regime. We validate this further with approximated models for heptameric~\cite{qian:scadv:2021, Richard:photor:2023} and octameric LH2 complexes~\cite{schulten:pnas:1998, koepke:structure:1996, MALLUS:2018:CP}. Importantly, utilizing the same logic we propose a stacked model ring of diameter 400 \textit{nm}, which estimates highly efficient inter-layer excitation transfer (from sparse to dense arrangements) at zero temperature via the most sub-radiant eigenmode. Recent experimental progresses~\cite{Browaeys:np:2014, Hemmig:nl:2016, Rueda:jpcc:2024} add to the feasibility of this nanostructure.

Following we provide a brief structure of this manuscript. In Section.~\ref{model}, we describe the model ring geometry under consideration and provide some theoretical descriptions together with Appendix-\ref{apen1}. In Section.~\ref{optimize-LH2}, we theoretically explore the possibility of optimizing the inter-layer excitation transfer with model biological LH2 geometries with $C_9$ symmetry, when all two-level emitters are photoactive at a single wavelength (some details are in Appendix-\ref{eigen-apen}). We then estimate the maximum excitation energy transfer for non-identical emitter layers, as present in the biological LH2 complexes with the free-space model, and provide some discussions (some details are provided in the Appendix-\ref{apen800-850}). Thereafter, we further validate our outcomes with model stacked rings with $C_7$, $C_8$ symmetry (technical details are in Appendix-\ref{apen-c7c8}). In Section.~\ref{energyflow}, we discuss the asymmetric energy flow in the layers of the model LH2 structure. In the following Section.~\ref{nanoscale}, we first discuss the theoretical strategy for designing the bio-mimicked nanoscale geometry, which exhibits efficient excitation transfer via the most sub-radiant mode in between the ring layers at zero temperature (some details are provided in Appendix-\ref{apen-nano}). To push our results towards a realistic experimental scenario, we discuss the effects of on-site static disorder and dephasing in the excitation transfer (also in Appendix-\ref{disorder}). In Section.~\ref{conclusions} we supply a summary and outlook of this research endeavor. Finally in the Section.~\ref{data} we provide the data availability statements.

\section{Model}
\label{model}
\begin{figure}[h]
\centering
\includegraphics[width=0.95\linewidth]{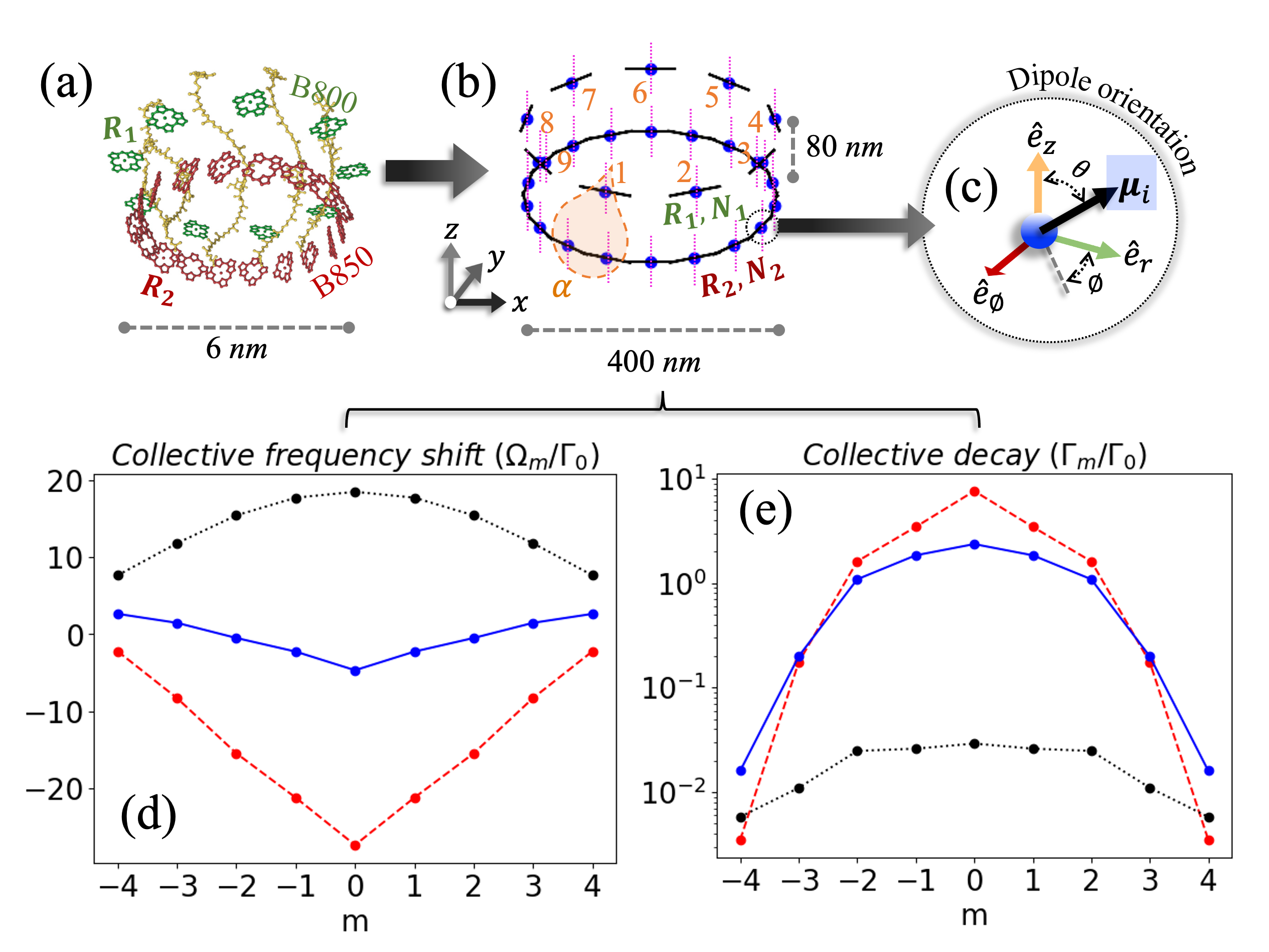}
\caption{(a) Pigment arrangements of biological LH2 complex of purple photosynthetic bacteria \textit{Rbl. acidophilus} (image taken from~\cite{bujak:APL:2011} and flipped vertically for the analogy) with bacteriochlorophyll (BChl) (Green: BChl B800, Red: BChl B850) and carotenoids (in yellow). The diameter of the B800 ring is around 6 \textit{nm} (parameters are taken from Ref.~\cite{Montemayor:JPCB:2018} and listed in Appendix-\ref{apen800-850}). (b) Bio-mimicked enlarged stacked concentric nanoring structure formed of two-level emitters with 800 \textit{nm} transition wavelength (parameters in Table-\ref{tab2}). We consider one BChl as a point-like emitter (blue circles) with fixed dipole orientation. Top ring ${R_1}$ is with $N_1 = 9$ emitters. Bottom ring ${R_2}$ has $N_2 = 18$ emitters. They are arranged in two concentric rings (${R_{2_a}}$ and ${R_{2_b}}$ with $N_{2_{a(b)}} = 9$ dipoles each). Small black-solid lines (or magenta-dotted lines) indicate the tangential (transverse) dipole orientation. The stacked geometry combines into nine unit cells. The orange-dashed contour `$\alpha$' indicates one unit cell, which contains $d = 3$ dipoles. (c) Angles ($\theta, \phi$) define the orientation of the $i^{th}$ dipole with dipole-moment $\boldsymbol{\mu}_i$. (d) Collective frequency shifts $(\Omega_m/\Gamma_0)$ and (e) decay rates $(\Gamma_m/\Gamma_0)$ for the eigenmodes with angular momentum $m$, corresponding to the geometry in (b). For each 800 \textit{nm} dipole the calculated spontaneous emission rate is $\Gamma_0 \sim$ 25.7 MHz.}
\label{fig1}
\end{figure}
We consider an LH2-inspired~\cite{McDermott:nature:1995} nonameric geometry of two concentric three-dimensional (3D) stacked ring layers of two-level quantum emitters. Ring ${R_i}$ contains $N_i$ dipoles, where $i=1,2$ (see Fig.~\ref{fig1}(b)). The emitters are identical (with transition frequency $\omega_0$) and separated by sub-wavelength distances, and thus, experiencing dipole-dipole interactions, i.e., exchange of excitation (for instance, $| e_i, g_j\rangle \leftrightarrow | g_i, e_j\rangle$). Here $| e_i(g_i) \rangle$ is the excited (ground) state of $i^{th}$ participating dipole. The interaction Hamiltonian ($H_{\rm DD}$) can be written as,
\begin{equation}
\label{Hdd}
    H_{\rm DD} = \sum_{i, j; i \neq j} \Omega_{ij} \hat{\sigma}^+_i \hat{\sigma}^-_j~. 
\end{equation}
Here $\hat{\sigma}^{+(-)}_i$ is the atomic raising (lowering) operator for the $i^{th}$ dipole. The collective dipole-dipole coupling $\Omega_{ij} = - (3\pi\Gamma_0/k_0){\boldsymbol{\mu}}^*_i \cdot {\rm Re}\left[ {\bf G}({\bf r}_{ij}, \omega_0)\right]\cdot {\boldsymbol{\mu}}_j$. Here ${\boldsymbol{\mu}}_i$ is the unit dipole orientation vector characterized by the angles $(\theta, \phi)$ (see Fig.~\ref{fig1}(c)), the spontaneous emission rate of one emitter is $\Gamma_0 = |\boldsymbol{\mu}|^2 k^3_0/(3\pi\hbar\epsilon_0)$ and ${\bf G}({\bf r}_{ij}, \omega_0)$ is the free space Green's tensor propagator (see Appendix-\ref{apen1}). The dipole-dipole interaction varies as a polynomial of the separation of inter-atomic distances and has the following form~\cite{FICEK:pr:2002},
\begin{table}[h!]
\caption{\label{tab2}Geometric parameters in use for LH2-inspired nanorings [Fig.~\ref{fig1}(b)]. Top ring is ${R_1}$ ($N_1 = 9$ emitters). Bottom ring ${R_2}$ consists of two rings : ${R_{2_a}}, { R_{2_b}}$ ($N_{2_{a(b)}} = 9$ emitters). Vertical layer separation is $Z_1$. All 27 dipoles are photoactive at 800 \textit{nm} and in the tangential orientation.}
\begin{ruledtabular}
\begin{tabular}{ccc}
{Ring radius ($r_i$) } & {Ring rotation} &  {Dipole orientations}\\
{Layer separation ($Z_1$)} &  $(\nu_i)$ &  $(\theta_i, \phi_i)$ \\
(in \textit{nm}) &  (in $deg$) &  (in $deg$)\\
\hline
${r_1}$ \hskip 0.5cm 200 &  $\nu_1$ \hskip 0.5cm $0^{\circ}$ &  $\theta_{1},\phi_1$ \hskip 0.7cm $90^{\circ}, 90^{\circ}$\\
$Z_{1}$ \hskip 0.6cm 80 &  & \\\hline
$r_{2_a}$ \hskip 0.5cm 200  &   $\nu_{2_a}$ \hskip 0.5cm $0^{\circ}$ &  $\theta_{2_a},\phi_{2_a}$ \hskip 0.5cm $90^{\circ}, 90^{\circ}$\\
$r_{2_b}$ \hskip 0.5cm 200 &   $\nu_{2_b}$ \hskip 0.4cm $20^{\circ}$ &  $\theta_{2_b},\phi_{2_b}$ \hskip 0.5cm $90^{\circ}, 90^{\circ}$ \\
\end{tabular}
\end{ruledtabular}
\end{table}
\begin{align}
\Omega_{ij} = \frac{3\Gamma_0}{4} \Bigl[(1- 3 \cos^2 \theta) &\left( \frac{\sin \xi_{ij}}{\xi^2_{ij}} + \frac{\cos\xi_{ij}}{\xi^3_{ij}}\right) \nonumber\\
&- \sin^2 \theta \frac{\sin\xi_{ij}}{\xi_{ij}} \Bigr]~,
\label{omegadd}
\end{align}
where $\xi_{ij} = k_0 r_{ij}$ and $\theta$ is the angle between dipole moment $\boldsymbol{\mu}$ and the separation vector of two dipoles ${\bf r}_{ij}$. Under the Born-Markov approximation~\cite{asenjo:prx:2017} the master equation ($\hbar =1$) is written as $\dot{\rho} = -i [H_{\rm DD}, \rho] + \mathcal{L} [\rho]$, where $\rho$ is the density matrix for the emitter degrees-of-freedom and $\mathcal{L} [\rho]$ is the Lindblad term, $\mathcal{L}[\rho] = \sum_{i,j} \Gamma_{ij}/2 \left(2\hat{\sigma}^-_j \rho \hat{\sigma}^+_i - \hat{\sigma}^+_i \hat{\sigma}^-_j\rho - \rho\hat{\sigma}^+_i \hat{\sigma}^-_j \right)$. Here the collective dissipation term is connected to the imaginary part of the Green's tensor as $\Gamma_{ij} = (6\pi\Gamma_0/k_0){\boldsymbol{\mu}}^*_i \cdot {\rm Im}\left[ {\bf G}({\bf r}_{ij}, \omega_0)\right]\cdot {\boldsymbol{\mu}}_j$. From now on we will limit ourselves to the single-excitation manifold, since we essentially deal with a single excitation or photon. This allows us to work with an effective non-Hermitian Hamiltonian of the following form~\cite{Mariona:pra:2019}, 
\bea
H_{\rm eff} = \sum_{i,j} \left(\Omega_{ij} - i \frac{\Gamma_{ij}}{2}\right) \hat{\sigma}^+_i \hat{\sigma}^-_j~.
\label{Heff}
\eea
Here, the on-site collective energy shifts are neglected ($\Omega_{ii} = 0$), as they lead to a simple renormalization of the bare transition frequency $\omega_0$. The $C_N$-symmetry of the ring geometry enables a Bloch eigenmode description with angular momentum quantum numbers $m$~\cite{Mariona:pra:2019}. For the geometry previously described and also depicted in Fig.~\ref{fig1}(b), we can re-express Eq.(\ref{Heff}) in the following form (see Appendix-\ref{apen1} for details),
\bea
H_{\rm eff} = \sum_{m} \sum_{\lambda \in \{1,2_a,2_b\}} \left(\Omega_{m\lambda} -i \frac{\Gamma_{m\lambda}}{2}\right) \hat{\sigma}^+_{m\lambda} \hat{\sigma}^-_{m\lambda}~.
\eea
In this case, for any given value of $m$ there are three possible solutions ($\lambda \in \{1,2_a,2_b\}$). Here the index $j$ for solutions, i.e., $j=(1,2_a,2_b)$ corresponds to the ring ${R_j}$. The collective energy shifts $\Omega_m$ and collective decay rates $\Gamma_m$ for a particular choice of the system parameters (indicated in Table-\ref{tab2}) are shown in Fig.~\ref{fig1}(d) and (e) with nonameric stacked ring of nanometer dimension.

\section{Optimizing excitation transfer between layers in an LH2 complex}
\label{optimize-LH2}
At a very small separation $|{\bf r}_{ij}|\ll \lambda$ the dipole-dipole interaction $\Omega_{ij}/(3\Gamma_0/4)$ [Eq.(\ref{omegadd})] predominantly varies as $\sim1/r^3_{ij}$, which is the case for biological LH2s. To calculate the inter-layer excitation transfer ${R_1}$ $\Rightarrow$ ${R_2}$, we take one definite eigenmode $m$ as an initial state, i.e., $|\Psi_m(t=0)\rangle$ from the sparse (top) ring ${R_1}$ [Fig.~\ref{fig1}(a)] and compute the total population transferred to the dense (bottom) ring(s) ${R_2} \equiv \{R_{2_a}\} \cup \{R_{2_b}\}$ at time $t$ with $|\Psi_m(t)\rangle$, i.e., 
\bea
\langle \hat{\sigma}^{ee}_{m} (t) \rangle_{R_2} = \sum_{i \in {R_{2_a}}} \langle \hat{\sigma}^{+}_{i} \hat{\sigma}^{-}_{i} \rangle_m + \sum_{j \in {R_{2_b}}} \langle \hat{\sigma}^{+}_{j} \hat{\sigma}^{-}_{j} \rangle_m~.
\label{exc-tran}
\eea
\begin{figure}[h]
\centering
\includegraphics[width=\linewidth]{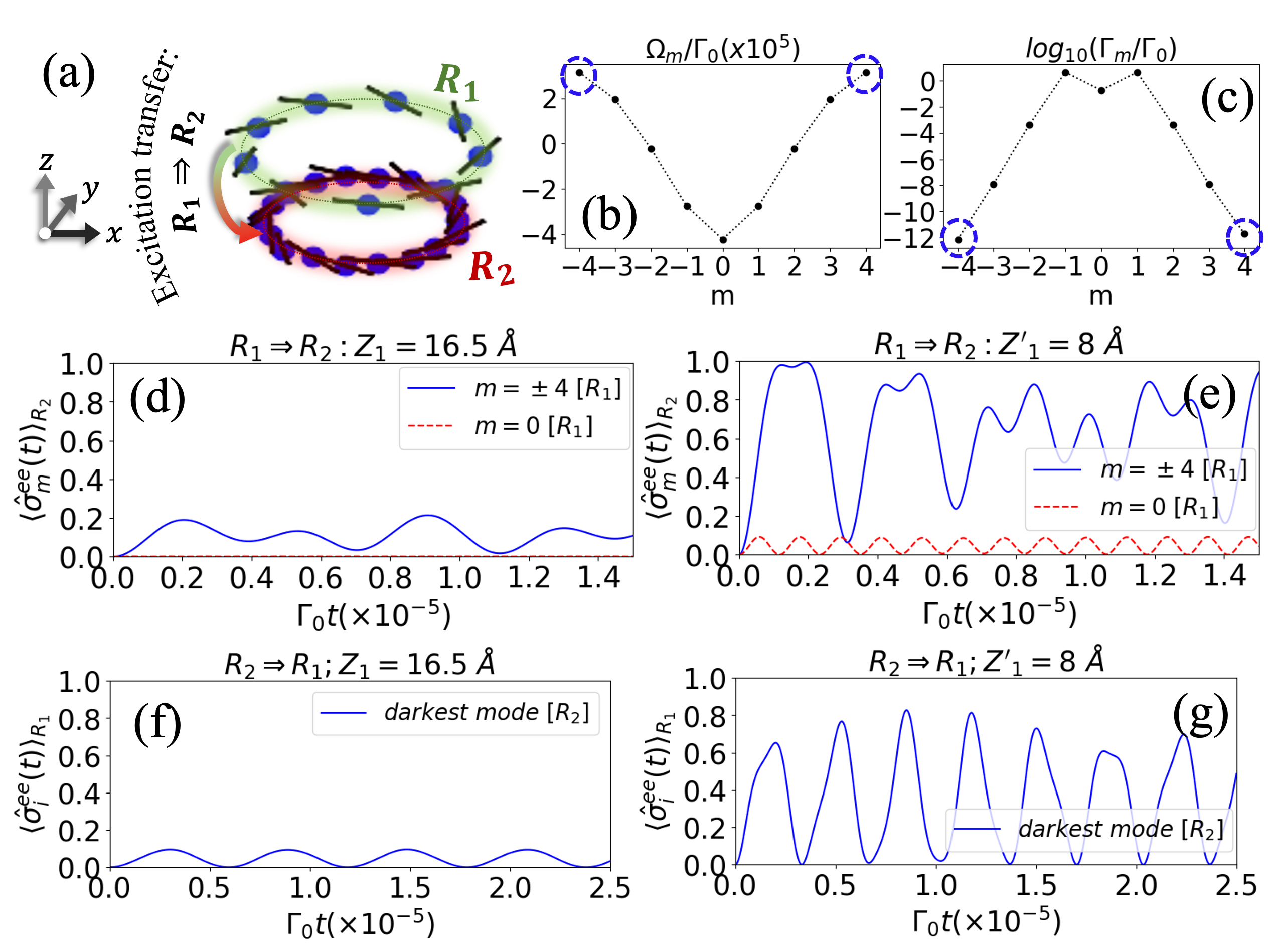}
\caption{(a) Bio-ring geometry (not to scale) with all 800 \textit{nm} dipoles (blue-solid circles) and biological dipole orientations (indicated by small black-solid lines). Parameters are taken from Ref.~\cite{Montemayor:JPCB:2018} and enlisted in Appendix-\ref{apen800-850}. (b) The collective energy shift ($\Omega_m/\Gamma_0$) and (c) collective decay rate ($\Gamma_m/\Gamma_0$) of the top-ring ${R_1}$ for different modes $m$. Blue-dotted circles denote the darkest mode $m=\pm 4$. Variation of $\langle\hat{\sigma}^{ee}_m (t)\rangle_{R_2}$ with scaled time $\Gamma_0t$ for $m=\pm 4$ (higher-transfer) and $0$ (less-transfer) at (d) $Z_1 = 16.5$ {\AA} (biological vertical layer separation \cite{Montemayor:JPCB:2018}) and (e) $Z'_1 = 8$ {\AA} ($Z'_1 < Z_1$). For $Z'_1$ the maximum value of inter-ring excitation transfer ${R_1}\Rightarrow{R_2}$, can be boosted to 100\% (e) from 21\% (d), for the most sub-radiant mode $m = \pm 4$ (blue-solid curves) of ring ${R_1}$. Plots (f) and (g) show the quantitative estimations for the reverse process ${R_2}\Rightarrow{R_1}$ for the darkest mode of ring ${R_2}$. Those displays around  10\% and 80\% transfer to ring ${R_1}$, respectively.}
\label{fig2}
\end{figure}
As a figure-of-merit of the efficiency of the inter-layer excitation transfer, we calculate ${\rm Max}[\langle \hat{\sigma}^{ee}_{m} (t) \rangle_{R_2}]$ for a time-bin $T\in \{0, \Delta_t\}$. Generally, the minimum value of $\Delta_t$ is guided solely by the ring size or the nearest-neighbor separation under consideration, i.e., a denser arrangement of dipoles - a faster and more sparse - a slower process. In Fig.~\ref{fig2}(d) and (e) we display the inter-layer excitation energy transfer for the symmetric $m=0$ and very dark $m=\pm 4$ eigenmodes [Fig.~\ref{fig2}(b), (c)], for the bio-geometry [Fig.~\ref{fig2}(a)]. The considered time bin is $\Delta_t \sim 0.6~ps$, i.e., extremely fast excitation transfer~\cite{sundstrom:jpcb:1999} at sub-nano dimension. As expected the darkest mode ($m = \pm 4$) assists in the highest inter-layer excitation energy transfer~\cite{Mariona:pra:2019}. For bio-geometry, we obtain a maximum of 21\% excitation transfer [Fig.~\ref{fig2}(d)]. However, with decreased vertical ring separation, i.e., at $Z_1 = $ 8 {\AA} we numerically obtain a maximum of 100\% excitation energy transfer [Fig.~\ref{fig2}(e)]. The key idea is to make the inter-layer nearest neighbor separation very close to the intra-ring nearest neighbor separation of dense ring ${R_2}$, i.e., $\sim$10, 11 {\AA}, where it is 17.6 {\AA}, 18.3 {\AA} in the biological LH2 (see Table-\ref{tab3}). This reforms one unit cell to be shaped like an equilateral triangle and therefore allows better hybridization for some of the eigenmodes (majorly the darker ones). This eventually assists in the efficient excitation energy transfer via the most sub-radiant mode in Fig.~\ref{fig2}(e). See Appendix-\ref{eigen-apen} for further details. In general, by considering the radius of all rings equal (this reshapes the inter-layer unit cell triangle), one can also boost the excitation transfer. Note that these proposed structural modifications are only a theoretical possibility at the sub-nano dimension.

\begin{table}[h]
\caption{\label{tab3}Estimated nearest-neighbor distances $(r_{ij})$ and inter-layer separations ($Z_1$) (in {\AA}) in the $C_9$ LH2 [Fig.~\ref{fig1}(a)] (from Refs.~\cite{Montemayor:JPCB:2018,Kohler:revbio:2006}), the approximated distances in LH2 structure for Fig.~\ref{fig2}(e) and LH2-inspired nanoring [Fig.~\ref{fig1}(b)].}
\begin{ruledtabular}
\begin{tabular}{lccc}
 Description & {LH2} &  {Modified LH2} &  {Nano rings}\\
\hline
Intra-ring ${R_1}$ ({\AA})& 21.3~\cite{Kohler:revbio:2006} &  $\sim$ 22 &  1368 \\
\hline
Ring ${R_{2_{a(b)}}}$ ({\AA}) &  9.2, 9.5~\cite{Kohler:revbio:2006} & $\sim$ 9.4 &  694 \\
\hline
Inter-ring ({\AA})& 17.6, 18.3~\cite{Kohler:revbio:2006} & $\sim$ 10.5, &  800, \\
(${R_1}$ - ${R_{2_{a(b)}}}$) & & 11.2&  1059  \\
\hline
$Z_1$({\AA})& 16.5~\cite{Montemayor:JPCB:2018} & 8 &  800 \\
\end{tabular}
\end{ruledtabular}
\end{table}

Nature's rings display complex oligomeric designs~\cite{koepke:structure:1996, Kohler:revbio:2006}. The presence of other important elements~\cite{Kohler:revbio:2006}, for example, $\alpha/\beta$-apoproteins, low-wavelength photo-active carotenoids, and an environment of different chemical interactions make these LH2 complexes quite intricate, yet highly efficient. In reality, these complexities perhaps prevent having a smaller vertical inter-layer separation in LH2. In biological LH2, nature relies on two circular layers of non-identical emitters (B800 and B850 bands). Our theoretical estimation shows around 37\% and 44\% inter-layer excitation transfer (higher than Fig.~\ref{fig2}(d)) for certain modes of sparse ring ${R_1}$ (see Appendix-\ref{apen800-850} for details) with biological LH2 complex. However, the concrete reason why the ring ${R_2}$ is exactly active at 850 \textit{nm}, might be the collective shift due to the dense dipole arrangements~\cite{schulten:2009:jcp} and perhaps some other reasoning coming from the complex LH2 structure, which is yet to be understood concretely. The physical benefit of considering 850 \textit{nm} dipoles in the bottom ring (denser arrangements of dipoles) results in the modification of eigen-energy bands~\cite{verena:nm:2023} which seems to enable increased excitation transfer via certain eigenmodes.

We further validate our discussed hypothesis with theoretical estimations of approximated models for heptameric~\cite{qian:scadv:2021, Richard:photor:2023} and octameric biological LH2~\cite{koepke:structure:1996,schulten:pnas:1998, MALLUS:2018:CP} of purple bacteria. We observe boosted excitation transfer for decreased vertical inter-layer separation in both cases considering all dipoles having transition wavelength of $800~nm$ (see Appendix-\ref{apen-c7c8} for details). In particular, with $C_7$ symmetry, we obtain a maximum of 99.7\% excitation transfer; for $C_8$, it is around 80\%. Note that the dipole orientations used for $C_7$ and $C_8$ cases are assumed to obtain an overview.

On a related note, for a nanoscale stacked ring, the inter-dipole separations will be larger than that of biological LH2. As a result, it would show much smaller collective energy shifts [Fig.~\ref{fig1}(d)] compared to the LH2 sub-nano rings (Appendix-\ref{apen800-850}). Thus consideration of all identical emitters would suffice the design requirement. This is evident in nature's rings with $C_7$ symmetry, where the increased inter-emitter distances cause a blue shift of the B850 band (as seen in the $C_9$ case), resulting in the B828 band for $C_7$ LH2 complex~\cite{qian:scadv:2021}.

\section{Asymmetric energy flow in the ring layers of LH2}
\label{energyflow}
The reverse excitation transfer process, i.e., ${R_2}\Rightarrow{R_1}$ is quantitatively dissimilar and expectantly reflects a small update in the time scale of variation (see Fig.~\ref{fig2}(f) and (g)). In Fig.~\ref{fig2}(f), (g) the darkest mode may correspond to $m=\pm 4$ mode of ${R_2}$. Thus, our results show that the excitation or energy transfer is more efficient from a \textit{sparse} to \textit{dense} ring layers, mainly via the darkest eigenmode in this cylindrical assembly of quantum emitters. Perhaps nature thus relies on sparse and dense circular self-aggregation of BChls in the oligomeric layers of LH2 complexes for efficient directional transfer of the collected solar energy. Note that our estimations are in the Markovian picture and the effects of thermal motion~\cite{pant:pccp:2020} are not taken into account. Intuitively it seems that this feature would be the reason for efficient funneling of energy in photosynthesis through the planar multi-ring conformation. For instance, LH2$\Leftrightarrow$LH2 (usually through the denser B850 ring) and then LH2 (18 BChl: B850)$\Rightarrow$LH1 (32 BChl: B875)~\cite{FLEMING:structure:1997}. For both cases the intra-ring dipole-dipole separations are around 9 {\AA}~\cite{Kohler:revbio:2006} and ring-radii ($r_i$) are different for LH1 and LH2 ($r_{B875} > r_{B850}$)~\cite{Kohler:revbio:2006}. In principle, this feature may possibly remains applicable to other sub-wavelength quantum emitter arrangements, although these claims demands further investigations in depth for concreteness.

\section{Efficient inter-layer excitation transfer in a stacked nanoring} 
\label{nanoscale}
Next, we consider the same geometry as before and attempt to scale it up in the nanoscale regime [Fig.~\ref{fig1}(b)]. Rings are formed by two-level quantum emitters, which are photoactive at 800 \textit{nm}. Equal diameters in the top and bottom ring and the chosen inter-layer separation (see Table-\ref{tab2}) enable the hybridization for some eigenmodes (see Appendix-\ref{apen-nano}). Unlike the previous geometries, we do not form a unit cell to be shaped like an equilateral triangle (see Table-\ref{tab3}). There are two inter-ring pairs: one with $Z_1 = 0.1 \lambda$ and another with a larger separation. The minimal intra-ring nearest-neighbor separation is approximately $\sim 0.08\lambda$~\cite{Hemmig:nl:2016, Rueda:jpcc:2024}. This principle should be applicable even for other sets of numbers, for instance, intra-ring dipole separation ($r_{ij}$) in ${R_2}$ and $Z_1$, both are equal $\sim 0.08 \lambda$ or $Z_1< r_{ij}$ in ${R_2})$ and could be used for designing even larger sub-wavelength rings, depending on experimental resources~\cite{Browaeys:np:2014, Hemmig:nl:2016,rui2020subradiant,lounis:nc:2022,lodhal:science:2023, Hood:np:2024, Rueda:jpcc:2024}. However, with the increase in ring size excitation transfer generally decreases, since the dipoles start to experience one another less strongly. 

\begin{figure}[h]
\centering
\includegraphics[width=\linewidth]{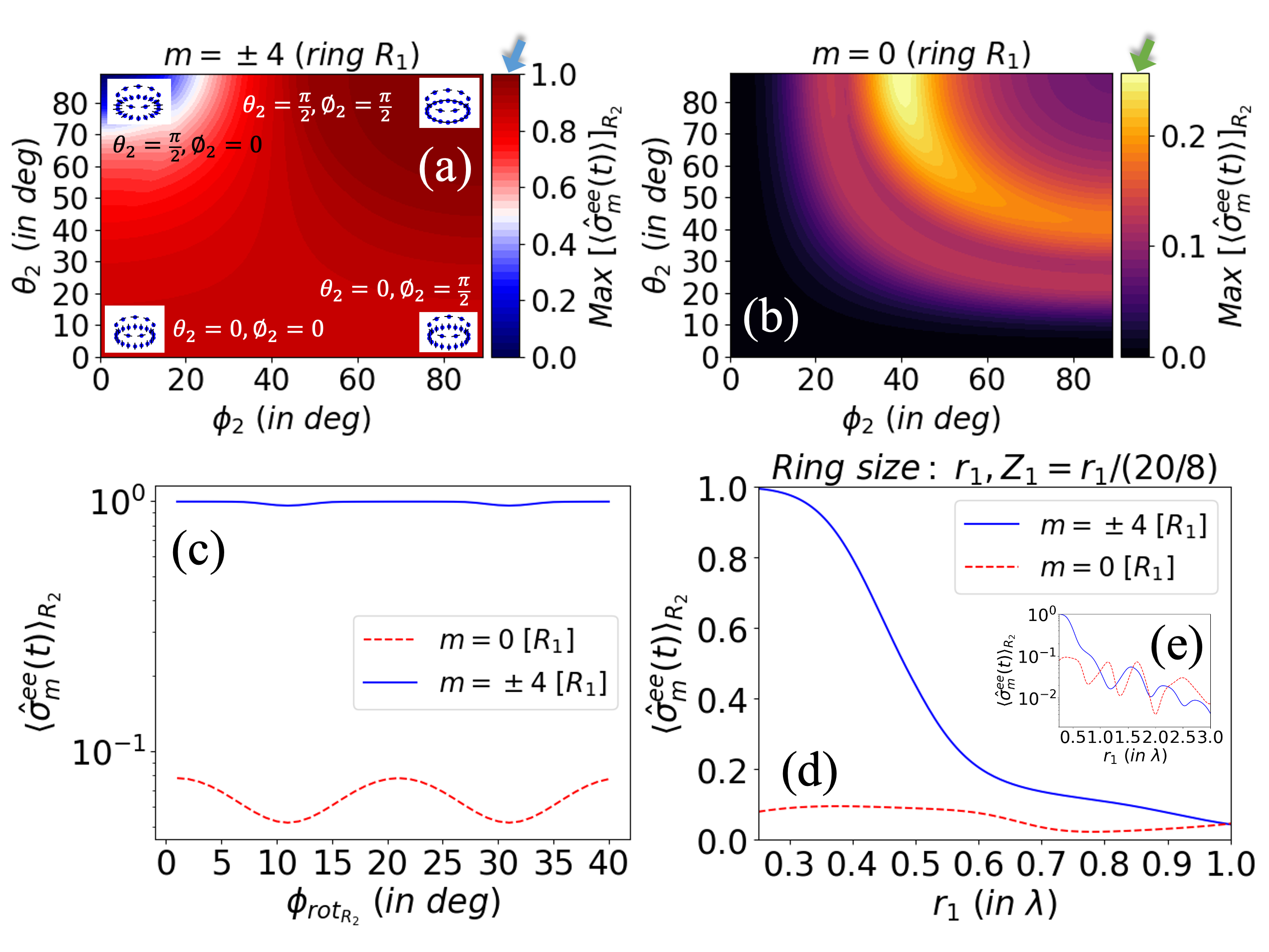}
\caption{Inter-layer excitation transfer (${R_1}\Rightarrow{R_2}$) for nanoring [Fig.~\ref{fig1}(b)]: with tunable dipole-orientations in $R_2$ (a)-(b), rotation of $R_2$ (c) and increase of ring-size (d). We consider the dipoles to be tangentially oriented in ring ${R_1}$. For different dipole orientations in ring ${R_2}$ (angles $\theta_2, \phi_2$), we show ${\rm Max}[\langle\hat{\sigma}^{ee}_m(t)\rangle_{R_2}] \in \{0,1\}$ for the anti-symmetric sub-radiant mode $m=\pm 4$ (a) and for the symmetric radiant mode $m=0$ (b) of sparse ring ${R_1}$. A few example snaps for dipole orientations in the ring layers are displayed for the mentioned ($\theta_2, \phi_2$) values in the plot (a). For certain dipole orientations in $R_2$ the $m=\pm 4$ exhibits a maximum of excitation transfer $\sim$ 99.5\%, i.e., ${\rm Max} (\langle\hat{\sigma}^{ee}_m(t)\rangle_{R_2}) \sim 1$ (indicated by the blue arrow). In contrast, for the radiant mode, $m=0$ the excitation transfer is generally less (indicated by the green arrow), around 25\% (b). We consider all dipoles to be in tangential orientation (in both layers) for plots (c) and (d). (c) With the rotation of ring ${R_2}$ ($\phi_{{rot}_2}$), we plot ${\rm Max} (\langle\hat{\sigma}^{ee}_m(t)\rangle_{R_2})$ for the above mentioned modes. Plot (d) shows the effect of increasing ring size (ring radius $r_1$ and vertical ring separation $Z_1 = r_1/(20/8)$ are changed together) on excitation transfer with $m=0, \pm 4$. The inset (e) shows the same variation (as in (d)), but for longer $r_1$, i.e., up to $3 \lambda$.}
\label{fig4}
\end{figure}
We consider an eigenmode from the sparse-ring ${R_1}$ [Fig.~\ref{fig1}(b)] and calculate Eq.(\ref{exc-tran}) to quantify the inter-layer (${R_1} \Rightarrow {R_2}$) excitation transfer with tunable dipole orientations in the dense ring ${R_2}$ (see Fig.~\ref{fig4}). The dipoles in ring ${R_1}$ are in tangential orientation, i.e., $(\theta_1,\phi_1) = (90^{\circ}, 90^{\circ})$. The collective energy shift ($\Omega_m$) and collective decay rates ($\Gamma_m$) for ${R_1}$ are detailed in Appendix-\ref{apen-nano}. For certain dipole orientations in ${R_2}$, one can achieve a maximum of $\sim$ 1 i.e., 100\% excitation transfer within the geometry for the darkest mode [Fig.~\ref{fig4}(a)]. For the bright mode of $R_1$, the maximum transfer seems much less [see Fig.~\ref{fig4}(b)].
Here, time-bin $\Delta_t$ is around $116~ns$ (3$\Gamma_0$t) (see Appendix-\ref{apen-nano}). The reverse transfer process $R_2\Rightarrow{R_1}$ is, in general, smaller, i.e., it is more efficient from \textit{sparse} to \textit{dense} arrangements in the stacked ring layers. Appendix-\ref{apen-nano} illustrates further details on this. Hence at zero temperature, we estimate a highly efficient excitation transfer in between layers of stacked nanoscale ring geometry. On a related note, for the biological dipole orientations in LH2~\cite{Montemayor:JPCB:2018} but with our nanoscale rings, the estimated maximum excitation transfer turns out to be 0.99 for the darkest mode $m=\pm 4$ and 0.09 for the bright mode $m=0$ of ring ${R_1}$.

In Fig.~\ref{fig4}(c) and (d), we consider all dipoles to be in tangential orientation. The maximum excitation transfer shows minor changes with the rotation of ring ${R_2}$ and is maximized at the parameters in Table-\ref{tab2} since the minimum inter-layer dipole separation is achieved around then. The variation is periodic with a 20$^{\circ}$ rotation of ${R_2}$ due to the system symmetry [Fig.~\ref{fig4}(c)]. In Fig.~\ref{fig4}(d), we show that as the ring size increases, with $r_2 = r_1$ and $Z_1 = r_1/(20/8)$, the maximum energy transfer decreases. Inset (e) displays some oscillations at the larger ring sizes, i.e., $r_1 > \lambda$. In the super-wavelength regime, the excitation transfer is minute, however, it seems that the layer stacking possibly modulates the collective optical properties~\cite{chang:arxiv:2024, shahmoon:arxiv:2024} and exhibits beating pattern.

To simulate the efficiency of the excitation transfer in a more realistic experimental scenario, we consider the onsite static position and frequency disorder as well as dephasing mechanisms~\cite{Plenio_2008, Alan:JCP:2008, Rebentrost_2009, Chin:Njp:2010} for the darkest and the brightest eigenmodes of the sparse ring. As expected, we observe signatures of noise-assisted excitation transport~\cite{Plenio_2008,Alan:JCP:2008,Chin:Njp:2010, plenio:cp:2013,plenio:NJP:2014,hauke:pra:2018,roos:prl:2019,harush:prr:2020,peter:prr:2024} in the layers of LH2-inspired stacked nanoring (see Appendix-\ref{disorder} for details). 

\section{Conclusions}
\label{conclusions}
We have presented here a theoretical analysis based on biological stacked ring layers in the LH2 complex, to understand and engineer the influence of structural modification on establishing highly efficient inter-layer excitation transfer. We have proposed a bio-mimicked nanoscale stacked concentric ring geometry exhibiting efficient excitation transfer at zero temperature. In principle, this sub-wavelength nanostructure may be useful for future applications across various platforms~\cite{barredo:nature:2018,trotta:prl:2018,manzeli:nrm:2017,wallraff:nc:2018,Browaeys:np:2020,kaufman:np:2021,kirchmair:np:2022,lounis:nc:2022,lodhal:science:2023,Hood:np:2024}. The preliminary insights obtained here for asymmetric excitation energy flow via the sub-radiant modes, i.e., whispering gallery modes \cite{asenjo:prx:2017} in ring layers, may open possibilities for crafting inter-node lossless links in quantum networks~\cite{reiserer:rmp:2022}. While our model seems yet to predict efficient excitation transfer at the super-wavelength limit~\cite{chang:arxiv:2024, shahmoon:arxiv:2024}, further engineering of this stacked geometry may offer opportunities to improve scalability in the future. We acknowledge that our model is simple and has scopes for improvement to obtain better quantitative estimations, which biological LH2 complexes perhaps deserve. Nevertheless, the principles and insights discussed here will remain useful for modeling and engineering other light-matter platforms, for example, biomimetic  geometries~\cite{klug:1999,aida:angew:2004, Amico:pra:2023, Rueda:jpcc:2024} and emitter arrays~\cite{barredo:nature:2018,Janne:PRA:2023,pichard:prap:2024} to achieve efficient energy transfer.

\section{Data Availability}
\label{data}
The simulations in this manuscript were performed utilizing the QuantumOptics.jl~\cite{KRAMER2018} and CollectiveSpins.jl~\cite{collectivespins} frameworks in the Julia programming language. The plots were prepared using the Matplotlib library~\cite{matplotlib}. The data that support the findings of this manuscript are openly available \cite{pal:2024}.

\section*{Acknowledgements} 
This research was funded in whole or in part by the Austrian Science Fund (FWF) 10.55776/ESP682. We also acknowledge the funding from the FWF project Forschungsgruppe FG 5. R.H. acknowledges funding by the Austrian Science Fund (FWF) 10.55776/W1259. M.M.-C. acknowledges funding from Grant No. PID2020-114626GB-I00 from the MICIN/AEI/10.13039/501100011033 (Government of Spain).

\appendix
\section{Theoretical description of collective Bloch eigenmodes in stacked rings}

\label{apen1}
The free space Green's tensor propagator ${\bf G}({\bf r}, \omega_0)$ acts on a unit dipole $\boldsymbol{\mu}$ as following,
\begin{align}
    {\bf G}({\bf r}, \omega_0)\cdot  \boldsymbol{\mu} &= \frac{e^{ik_0r}}{4\pi r} \Bigl[(\boldsymbol{r}\times\boldsymbol{\mu}) \times \boldsymbol{r} + \nonumber\\
    & \left(\frac{1}{k^2_0 r^2}
    - \frac{i}{k_0 r}\right) (3 \boldsymbol{r}({\bf r}.\boldsymbol{\mu}) - \boldsymbol{\mu})\Bigr]~.
\end{align}
Here, $\boldsymbol{r}$ is the unit inter-particle separation vector, $k_0 = \omega_0/c = 2\pi/\lambda$ is the wave number for the specific emitter transition, and $\lambda$ is the corresponding transition wavelength. In the model considered in this work, where there are several emitters within a unit cell, the  effective Hamiltonian can be written as~\cite{asenjo:prx:2017},
\begin{align}
    H_{\rm eff}  =& - \mu_0 \omega_0^2\times\nonumber\\
    \sum^{N}_{i,j = 1} &\sum^d_{\alpha,\beta = 1} \left(\boldsymbol{\mu}^*_{i\alpha}\cdot {\bf G}({\bf r}_{i\alpha} - {\bf r}_{j\beta}, \omega_0)\cdot \boldsymbol{\mu}_{j\beta}\right)~ \hat{\sigma}^+_{i\alpha} \hat{\sigma}^-_{j\beta} \nonumber\\
    &\equiv \sum^{N}_{i,j = 1} \sum^d_{\alpha,\beta = 1} \mathcal{G}^{\alpha \beta}_{ij} \hat{\sigma}^+_{i\alpha} \hat{\sigma}^-_{j\beta}~.
\label{heffm}
\end{align}
Here $\mu_0$ is the vacuum permeability. Index $i(j)$ runs over the $N$ different unit cells, and $\alpha, \beta$ is an index denoting the different components in each cell. Each cell contains $d$ dipoles (in the studied model $d = 3$). The unit dipole orientation vector for $\alpha^{th}$ dipole in $i^{th}$ cell is $\boldsymbol{\mu}_{i\alpha}$.

In a rotationally symmetric ring, it can be shown~\cite{verena:nm:2023} that the two-site interaction terms only depend on the inter-site distance, such that 
\bea
\mathcal{G}^{\alpha \beta}_{ij} = \mathcal{G}^{\alpha \beta}_{i+1, j+1} \equiv \mathcal{G}^{\alpha \beta}_{l}~,
\eea
where $l$ is the site distance between unit cells, i.e., $l = j - i$ and $l = 0,1,..., N-1$. This allows us to rewrite the effective Hamiltonian as follows,
\bea
H_{\rm eff} = \sum_{i} \sum_l \sum_{\alpha,\beta} \mathcal{G}^{\alpha \beta}_{l} \hat{\sigma}^+_{i\alpha} \hat{\sigma}^-_{i+l,\beta}~.
\label{Heff:m}
\eea
Next, we can re-express the onsite operators by inverse Fourier transform as,
\bea
\hat{\sigma}^{+(-)}_{j\alpha} = \frac{1}{\sqrt{N}}\sum_m e^{-\mathrm{i} 2\pi m j/N} \hat{\sigma}^{+(-)}_{m\alpha}~.
\label{Eq:Heff_l}
\eea
Here, the orbital angular momentum quantum number is $m = 0, \pm 1, \pm 2, ..., \lceil \pm (N-1)/2 \rceil$, where $\lceil .\rceil$ is the ceiling function. By replacing these expressions into Eq.(\ref{Heff:m}), we arrive at~\cite{verena:nm:2023}, 
\bea
H_{\rm eff} = \sum_{m} \sum^d_{\alpha,\beta = 1} \tilde{\mathcal{G}}^{\alpha \beta}_m \hat{\sigma}^+_{m\alpha} \hat{\sigma}^-_{m\beta}~,
\label{Eq:Heff_m}
\eea
with 
\bea
\tilde{\mathcal{G}}^{\alpha \beta}_m = \sum^{N-1}_{l = 0} e^{i 2\pi m l /N } \mathcal{G}^{\alpha \beta}_{l}~.
\eea
The Hamiltonian in Eq.(\ref{Eq:Heff_m}) is already diagonal in the orbital degree of freedom associated with the unit cell index. For each of the quantum numbers $m$, we can further diagonalize the $d \times d$ complex matrix  $\tilde{G}^{\alpha \beta}_m$ and find the collective eigenmodes and corresponding eigenvalues, the real and imaginary part of which will correspond to collective frequency shift $(\Omega_{m\lambda})$ and collective decay rate $(\Gamma_{m\lambda})$. The effective Hamiltonian can be finally written as,
\bea
H_{\rm eff} = \sum_{m} \sum_{\lambda} \left(\Omega_{m\lambda} -i \frac{\Gamma_{m\lambda}}{2}\right) \hat{\sigma}^+_{m\lambda} \hat{\sigma}^-_{m\lambda}~, 
\eea
where $\hat{\sigma}^{+(-)}_{m\lambda}$ is the creation(annihilation) operator for the collective Bloch-mode of angular momentum $m$ and $\lambda \in \{1,2_a,2_b\}$ illustrates three possible solutions (for the studied model).


\section{Improved hybridization of eigenstates by altering the LH2 layer separation}
\label{eigen-apen}
In the biological nonameric LH2 \cite{McDermott:nature:1995}, the inter-layer unit cell is essentially a triangle with unequal sides. Roughly the inter-layer dipole-dipole separation is twice ($\sim$ 18 {\AA}) of the intra-ring nearest neighbor separation in the denser layer ($\sim$ 9 {\AA})~\cite{Kohler:revbio:2006}. Thus one can consider it to be an isosceles triangle. The dipole-dipole interaction $\Omega_{ij}/(3\Gamma_0/4)$ predominantly varies as $\sim 1/r^3_{ij}$ at sub-nano dimension. Here we consider one unit cell only and for simplicity, consider dipoles to be transversely orientated to the plane of the triangle, i.e., $\boldsymbol{\mu}_{i(j)}.\boldsymbol{r}_{ij} = 0$. If the intra-ring collective dipole-dipole coupling in the ring ${R_2}$ for the immediate vicinity is $\Omega(r)$, then inter-layer dipole-dipole coupling would be $\Omega/x^3$ ($x \sim 2$ for biological ring design and it is theoretically tunable). We define the bare states in the single-excitation manifold of the unit cell triangle of dipoles as follows,
\bea
\mid 1 \rangle &=& \mid e_{R_1}, g_{R_2}, g_{R_2} \rangle~,\nonumber\\
\mid 2 \rangle &=& \mid g_{R_1},e_{R_2}, g_{R_2} \rangle~,\nonumber\\
\mid 3 \rangle &=& \mid g_{R_1}, g_{R_2}, e_{R_2}\rangle~.
\label{barestate}
\eea
For an equilateral triangle, $x = 1$, and the eigenvalues are displayed in Table-\ref{tab7}. For $x>1$ the eigenvalues and corresponding eigenvectors are noted in Table-\ref{tab6}. To structure it similar to bio-geometry one should consider $x = 2$ (see Table-\ref{tab3} for approximated nearest-neighbour separations in different cases).
\begin{table}[h]
\caption{\label{tab7}
Eigenvalues ($E_{v_i}$) (scaled with $3\Gamma_0/4$) and eigenvectors $(\mid v_i \rangle)$ for a unit cell shaped like an equilateral triangle of sub-nano dimension.}
\begin{ruledtabular}
\begin{tabular}{lc}
{Eigenvalues ($E_{v_i}$)} & {Eigenvectors $(\mid v_i \rangle)$} \\
\hline
{$2\Omega(r)$} & $\frac{1}{\sqrt{3}}\left(\mid 1 \rangle + \mid 2 \rangle + \mid 3 \rangle \right)$\\
$-\Omega(r)$ &  $\frac{1}{\sqrt{2}}\left(-\mid 1 \rangle + \mid 3 \rangle \right)$\\
$-\Omega(r)$ &  $\frac{1}{\sqrt{2}}\left(-\mid 1 \rangle + \mid 2 \rangle \right)$\\
\end{tabular}
\end{ruledtabular}
\end{table}

\begin{table}[h!]
\caption{\label{tab6}
Eigenvalues ($E_{v_i}$) and eigenvectors $(\mid v_i \rangle)$ for a triangle where inter-dipole separation $|r_{ij}|$$\equiv r <<\lambda$ and dipole-dipole interaction $\Omega(r) \sim 1/r^3$ (scaled with $(3\Gamma_0/4)$). Here $C_1(x) = x^3 -\sqrt{x^6+8}$, i.e. $<0$; $C_2(x) = x^3+\sqrt{x^6+8}$, i.e., $>0$ and $x>1$.}
\begin{ruledtabular}
\begin{tabular}{lc}
{Eigenvalues ($E_{v_i}$)} & {Eigenvectors $(\mid v_i \rangle)$} \\
\hline
-{$\Omega(r)$} & $\frac{1}{\sqrt{2}} \left(-\mid 2\rangle + \mid 3 \rangle \right)$\\
$\frac{C_1(x)}{2x^3}\Omega(r)$ &  $\frac{1}{\sqrt{2+|C_2(x)|^2}} \left(-\frac{1}{2}C_2(x)\mid 1\rangle + \mid 2\rangle +\mid 3\rangle\right)$\\
$\frac{C_2(x)}{2x^3}\Omega(r)$ & $\frac{1}{\sqrt{2+|C_1(x)|^2}} \left(-\frac{1}{2}C_1(x)\mid 1\rangle  + \mid 2\rangle +\mid 3\rangle\right)$\\
\end{tabular}
\end{ruledtabular}
\end{table}
\begin{figure}[b]
\centering
\includegraphics[width=0.95\linewidth]{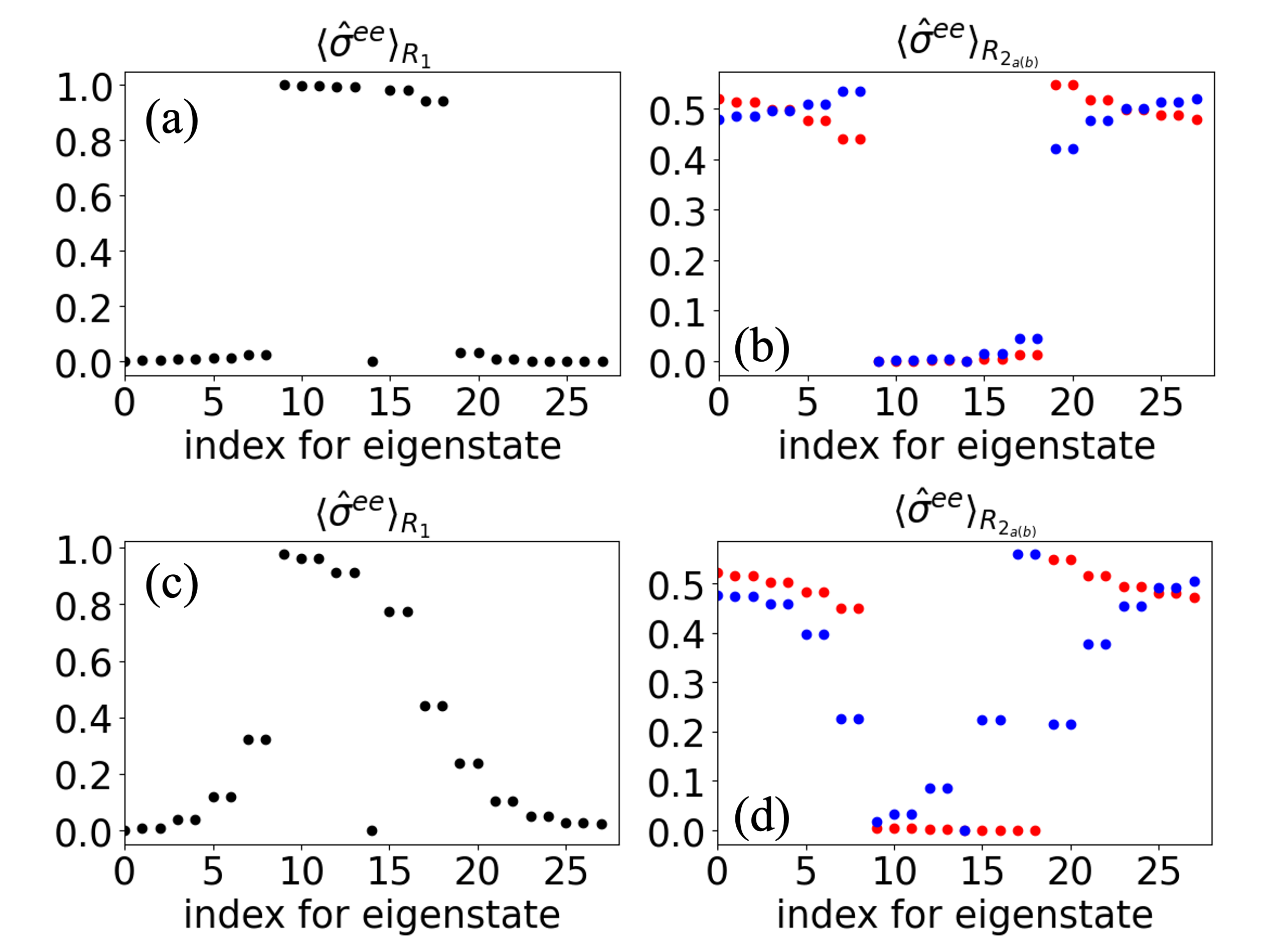}
\caption{Computed populations $\langle \hat n \rangle_i$ for the $i^{th}$ eigenstate of stacked LH2 rings. 28 eigenstates are indexed in $x$-axis. Plots (a) and (b) are with actual LH2 ring-layer separations, i.e., 16.5 {\AA} and bio-dipole orientations. Plots (c),(d) with the above dipole orientation, but with decreased vertical inter-layer separation 8 {\AA}; for all three rings, i.e., ${R_1}$ (in black), ${R_{2_a}}$ (in red) and ${R_{2_b}}$ (in blue), as displayed. In (c) and (d) some of the eigenstates (in ring $R_1$ and $R_3$ mostly, due to similar dipole orientations) exhibit shared populations in higher proportions, i.e., improved hybridization than the former case as in (a)-(b).}
\label{fig2a}
\end{figure}
From the eigenstates of one unit cell in Table-\ref{tab7} and \ref{tab6}, it is clear that the sub-radiant states have a higher contribution when $x=1$ (modified-geometry), than $x=2$ (biological spacings) from the bare-state $\mid 1\rangle$ [Eq.(\ref{barestate})], which carries the excitation of the top ring ${R_1}$.

Next, to prove this with LH2 ring layers with biological dipole orientations, we numerically compute populations, i.e. $\langle \hat n \rangle_i$ for the $i^{th}$ eigenstate of the geometry with original and also with decreased LH2 layer separation (in this case all other parameters are kept intact), respectively. All dipoles are considered to be photoactive at 800 \textit{nm} light. Here we would have $N_1+N_{2_a}+N_{2_b}+1 = 28$ eigenstates for the single excitation manifold. Fig.~\ref{fig2a}(a),(b) is for the biological LH2 layer separation (16.5 {\AA}) and Fig.~\ref{fig2a} (c),(d) for the decreased layer separation (8 {\AA}). The nearest neighbor separations for biological LH2 and modified LH2 are noted in Table-\ref{tab3} for comparison. We can immediately witness that some of the eigenmodes, in particular the darker ones, show shared populations in rings with higher proportions. This indicates the improved hybridization of the eigenstates of the rings. In particular, only $R_1$ and $R_{2_b}$ are highly hybridized since the dipoles in these rings have similar orientations (see Table-\ref{tab1}).


\section{Inter-layer excitation energy transfer in model nonameric LH2 rings}
\label{apen800-850}

We theoretically consider the 3D stacked geometry of the LH2 complex of \textit{Rbl. acidophilus} formed of two stacked circular layers of BChls, photoactive at 800 \textit{nm} and 850 \textit{nm} light, respectively \cite{McDermott:nature:1995}. We take the parameters from Ref.~\cite{Montemayor:JPCB:2018} (see Table-\ref{tab1}) and consider each bacteriochlorophyll (BChl) as one dipole. Each dipole is a two-level emitter. In particular, top-ring ${R_1}$ is formed with 800 \textit{nm} dipoles (sparse arrangement) and bottom-ring ${R_2}$ (formed of two ring ${R_1}$ and ${R_2}$) has all 850 \textit{nm} dipoles (dense arrangement). The dipole moments are $|\boldsymbol{\mu}_1| = 6.48$ D, $|\boldsymbol{\mu}_{2_a}| = 6.41$ D and $| \boldsymbol{\mu}_{2_b}| = 6.3$ D for rings, ${R_1}, {R_{2_a}}$ and ${R_{2_b}}$ respectively~\cite{Montemayor:JPCB:2018}. Dipole orientations are noted in Table-\ref{tab1} and obtained from Ref.~\cite{Montemayor:JPCB:2018}. 
\begin{table}[h]
\caption{\label{tab1} Geometric parameters for $C_9$ LH2 complex of \textit{Rbl. acidophilus} and taken from~\cite{Montemayor:JPCB:2018}. We denote the B800 ring as ring ${R_1}$ ($N_1 = 9$ emitters) and the B850 as ring 2 (${R_2}$), which consists of two sub-rings: ${R_{2_a}}, {R_{2_b}}$ ($N_{2_{a(b)}} = 9$ emitters). The ring radii are $r_1$, $r_{2_a}, r_{2_b}$ for rings $R_1, R_{2_a}$ and $R_{2_b}$, respectively. The vertical layer separation is $Z_1$. }
\begin{ruledtabular}
\begin{tabular}{ccc}
Ring size & {Ring rotation} &  {Dipole orientations}\\
($r_i$,$Z_1$) (in {\AA})&  $(\nu_i)$ (in $deg$) &  $(\theta_i, \phi_i)$ (in $deg$) \\
\hline
${r_1}$ \hskip 0.5cm 32.1 &  $\nu_1$ \hskip 0.3cm $23.3^{\circ}$ &  $\theta_{1},\phi_1$ \hskip 0.4cm $98.2^{\circ}, 63.7^{\circ}$\\
$Z_{1}$ \hskip 0.5cm 16.5 &  & \\\hline
$r_{2_a}$ \hskip 0.5cm 26  &   $\nu_{2_a}$ \hskip 0.3cm $-10.2^{\circ}$ &  $\theta_{2_a},\phi_{2_a}$ \hskip 0.2cm $96.5^{\circ}, -106.6^{\circ}$\\
$r_{2_b}$ \hskip 0.3cm 27.5 &   $\nu_{2_b}$ \hskip 0.4cm $10.2^{\circ}$ &  $\theta_{2_b},\phi_{2_b}$ \hskip 0.4cm $97.3^{\circ}, 60.0^{\circ}$ \\
\end{tabular}
\end{ruledtabular}
\end{table}

\begin{figure}[h]
\centering
\includegraphics[width=\linewidth]{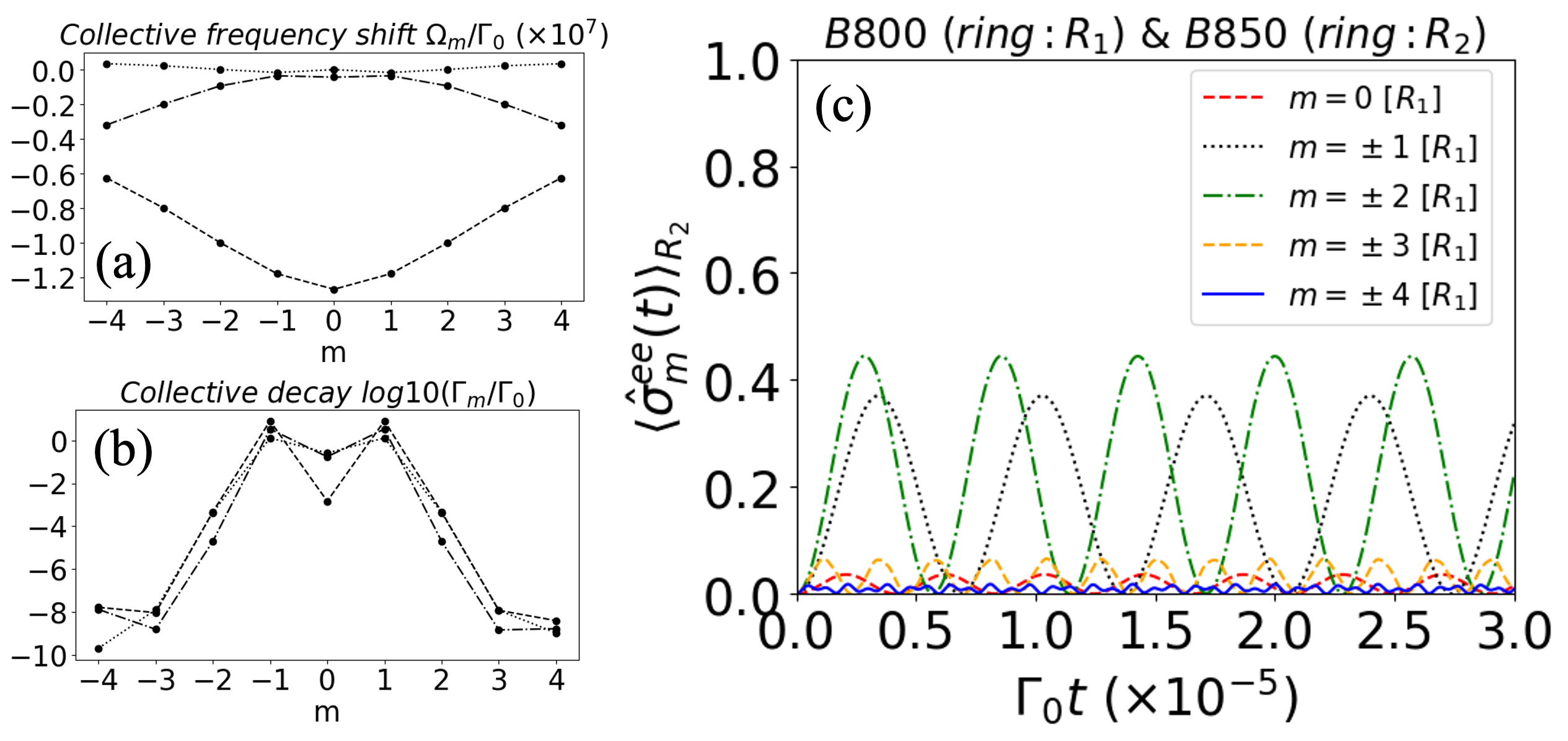}
\caption{Theoretical estimations with B800-B850 LH2 model $C_9$ rings (see Table-\ref{tab1} for parameters). (a) Collective energy shift ($\Omega_m/\Gamma_0$) and (b) collective decay rate ($\Gamma_m/\Gamma_0$) for angular momentum modes $m$. $\Gamma_0$ corresponds to the spontaneous emission rate for 800 \textit{nm} dipole $ \sim 25.7$ MHz. (c) Temporal evolution of the excitation energy transfer ${R_1}$ $\Rightarrow$ ${R_2}$ for different eigenmode $m$ of ring ${R_1}$ with scaled time $\Gamma_0 t$ (here, $3\times 10^{-5}\Gamma_0 t \equiv 1.1~ps$).}
\label{fig7}
\end{figure}

The total Hamiltonian for these interacting dipoles reads as,
\begin{equation} 
H_{\rm tot} = \sum_{i\in {R_1}} \omega_{800} \hat{\sigma}^+_i \hat{\sigma}^-_i +  \sum_{\substack{j \in {R_{2_a}}\\
j \in {R_{2_b}}}} 
\omega_{850} \hat{\sigma}^+_j \hat{\sigma}^-_j + H_{\rm DD}~,
\end{equation}
\begin{table*}[!ht]
\caption{\label{tab4} Approximated geometric parameters and dipole-orientations (assumed) for model heptameric LH2 of \textit{Mch. purpuratum}~\cite{qian:scadv:2021} (I) and octameric LH2 of \textit{Rsp. molischianum}~\cite{koepke:structure:1996,schulten:pnas:1998} (II). Calculated and reported [in parenthesis] intra- and inter-layer nearest neighbor separations are noted for comparison.}
\begin{ruledtabular}
I. \textit{Mch. purpuratum}\\
\begin{tabular}{cccc}
Ring size & Dipole orientations &  Ring rotation & Dipole-dipole separation \\
($r_i$, $Z_i$) ({\AA}) & ($\theta_i,\phi_i$) ($deg$) &  $\nu_i$ ($deg$) &  ({\AA}) \\
\hline
$r_1$\hspace{0.5cm} 28.5 & 98.2$^{\circ}$, 63.7$^{\circ}$ & 0$^{\circ}$ & 24.7 [24.2~\cite{qian:scadv:2021}] (B800 intra-ring) \\
$Z_1$\hspace{0.5cm} 18 & & & \\
\hline
$r_{2_a}$\hspace{0.5cm} 25.4 & 96.5$^{\circ}$, -106.6$^{\circ}$ &-13$^{\circ}$ & 22, 22.5 (B828 intra-ring)\\
$r_{2_b}$\hspace{0.5cm} 26 & 97.3$^{\circ}$, 60.0$^{\circ}$ &13$^{\circ}$ & 11.5 [9.6, 10.7~\cite{qian:scadv:2021}] (B828 intra-ring)\\
 & & & 19.1, 19.2, 25.4 [18.3, 19.9, 25.6~\cite{qian:scadv:2021}] (B800-B828 inter-ring)\\
\end{tabular}
\end{ruledtabular}
\begin{ruledtabular}
\\II. \textit{Rsp. molischianum}\\
\begin{tabular}{cccc}
Ring size & Dipole orientations &  Ring rotation & Dipole-dipole separation \\
($r_i$, $Z_i$) ({\AA})& ($\theta_i,\phi_i$) ($deg$) &  $\nu_i$ ($deg$)&  ({\AA}) \\
\hline
$r_1$\hspace{0.5cm} 29.2& 99$^{\circ}$, 60.8$^{\circ}$ & 0$^{\circ}$ &  22.3 [22~\cite{schulten:pnas:1998}, 22.3~\cite{MALLUS:2018:CP}] (B800 intra-ring) \\
$Z_1$\hspace{0.5cm} 18& &  & \\
\hline
$r_{2_a}$\hspace{0.5cm} 23& 93.7$^{\circ}$, -107.1$^{\circ}$&-12$^{\circ}$ &  18.3, 17.6 (B850 intra-ring) \\
$r_{2_b}$ \hspace{0.5cm} 24 & 98.9$^{\circ}$, 55.4$^{\circ}$ &12$^{\circ}$ & 9.8 [8.9, 9.2~\cite{schulten:pnas:1998}; 8.7, 9.9~\cite{MALLUS:2018:CP}] (B850 intra-ring)\\
 & & &  19.5, 19.8, 24.1 [19.4, 20.5~\cite{MALLUS:2018:CP}] (B800-B850 inter-ring)\\
\end{tabular}
\end{ruledtabular}
\end{table*}
where $\omega_{800(850)}$ is the atomic transition frequency of $800(850)~nm$ dipoles for respective rings. $H_{\rm DD}$ is the dipole-dipole interaction [Eq.(\ref{Hdd})]. The energy bands are shown in Fig.~\ref{fig7}(a) and collective decay in Fig.~\ref{fig7}(b) (also reported in Ref.~\cite{verena:nm:2023}). In Fig.~\ref{fig7}(b) the modes around the corners appear to be very dark and deviation from the expected decreasing feature seems to be a computational artifact. At biological ring sizes, i.e., $r_{ij} \ll \lambda$, the collective energy shift ($\Omega_{ij}$) (shown in Fig.~\ref{fig7}(a)) is quite large. The estimated maximum of inter-layer excitation transfer, i.e., ${\rm Max}[\langle\hat{\sigma}^{ee}_m(t)\rangle_{R_2}]$ for modes $m=\pm 1, \pm 2$ of $R_1$ is found to be around 37\% and 44\% (see Fig.~\ref{fig7}(c)). For the most sub-radiant mode $m=\pm 4$, unfortunately, the transfer is significantly less since the energy bands are quite far [Fig.~\ref{fig7}(a)]. In summary, the consideration of non-identical emitter layers at the sub-nano dimension results in the modification of the energy bands ($\Omega_{ij}$), which is crucial for achieving efficient excitation energy transfer through certain angular momentum modes of the running spin wave.

\section{Inter-layer excitation transfer in model heptameric and octameric LH2 rings}

\label{apen-c7c8}
Here we will show enhanced inter-layer excitation energy transfer using simplified coupled-dipole models of heptameric \cite{qian:scadv:2021} and octameric \cite{koepke:structure:1996,schulten:pnas:1998, MALLUS:2018:CP} single LH2 complexes, which have similar stacked layers of BChls as in LH2 complex of \textit{Rbl. aciodophilus} \cite{McDermott:nature:1995}, but exhibit $C_7$ and $C_8$ symmetry, respectively.

\subsection{Heptameric LH2}
\begin{figure}[h]
\centering
\includegraphics[width=\linewidth]{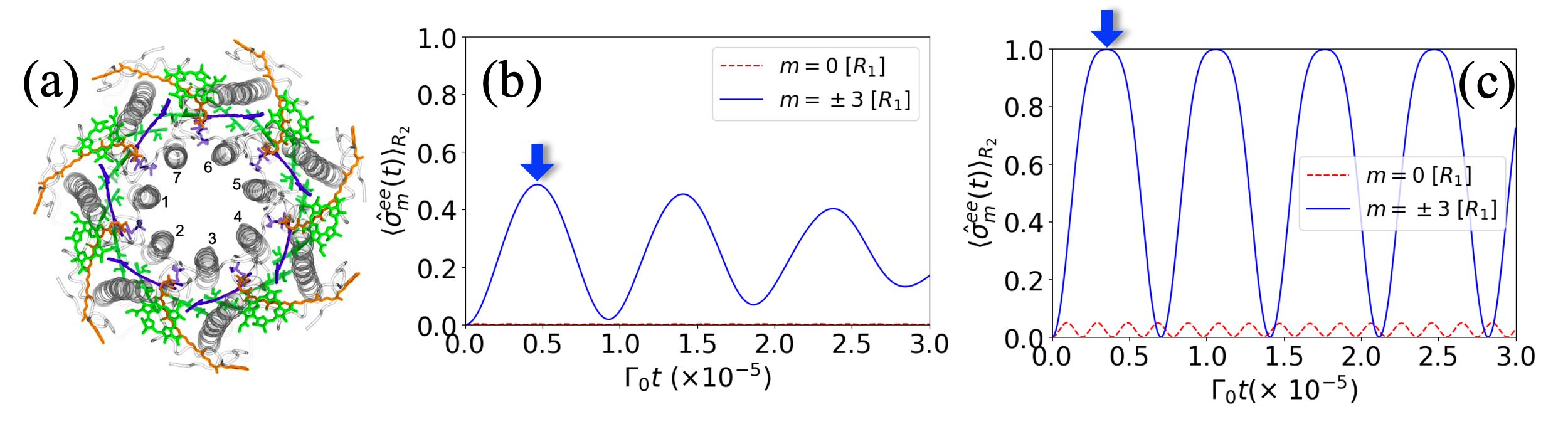}
\caption{(a) Structure of heptameric LH2 of \textit{Mch. purpuratum} (top view, taken from~\cite{Richard:photor:2023}). The prominent seven units are numbered, with each unit containing three BChls. BChls are shown in green, with the top layer containing 800 \textit{nm} BChls and the bottom layer containing 828 \textit{nm} BChls. The geometric parameters are listed in Table-\ref{tab4}(I). The $C_7$ symmetry supports seven eigenmodes. (b) The inter-layer excitation energy transfer (${R_1} \Rightarrow {R_2}$) for bio-layer separation estimates a maximum around $\sim 48.7\%$ (indicated by blue-arrow)), considering all dipoles to be photo-active at 800 \textit{nm} ($\Gamma_0 \sim 25.7$ MHz). (c) With a decreased vertical separation of $Z_1 = 9$ {\AA}, the maximum transfer is enhanced to approximately 100\% (indicated by the blue arrow), keeping other parameters unchanged.}
\label{800-828}
\end{figure}

The LH2 complex of the marine purple bacterium \textit{Marichromatium (Mch.) purpuratum} exhibits seven-fold symmetry (B800-B828)~\cite{qian:scadv:2021} (see Fig.~\ref{800-828}(a)). We use the nearest-neighbor separations reported in Ref.~\cite{qian:scadv:2021} for our analysis. For dipole orientations, we assume they are similar to those in \textit{Rbl. acidophilus} LH2 as described in Ref.~\cite{Montemayor:JPCB:2018}, with all dipole moments set to the same value, $|\boldsymbol{\mu}_{1}| = |\boldsymbol{\mu}_{2_a}| = |\boldsymbol{\mu}_{2_b}| = 6.48 \, \text{D}$. We extract BChl-BChl separations (both intra- and inter-layer) from Ref.~\cite{qian:scadv:2021} and calculate the approximate ring-size parameters (see Table-\ref{tab4}) for modeling the B800-B828 stacked rings.
In Fig.~\ref{800-828}, we present observations similar to those made for the $C_9$ case, assuming all dipoles are photoactive at 800 \textit{nm}. By decreasing the vertical separation from 18 Å to 9 Å, we theoretically achieve a unit cell shaped like an approximate equilateral triangle, with inter-ring side lengths of $\sim$11.4 Å, 10.7 Å, and an intra-ring length of $\sim$11.5 Å [Table-\ref{tab4}(I)]. Therefore this would allow better hybridization of the eigenmodes and facilitate enhanced inter-layer excitation transfer ${R_1} \Rightarrow {R_2}$. Specifically, the maximum of excitation transfer improves from 48.5\% [Fig.~\ref{800-828}(b)] to 100\% [Fig.~\ref{800-828}(c)] for the darkest eigenmode $m= \pm 3$ of ${R_1}$.

\subsection{Octameric LH2}
\begin{figure}[h]
\centering
\includegraphics[width=\linewidth]{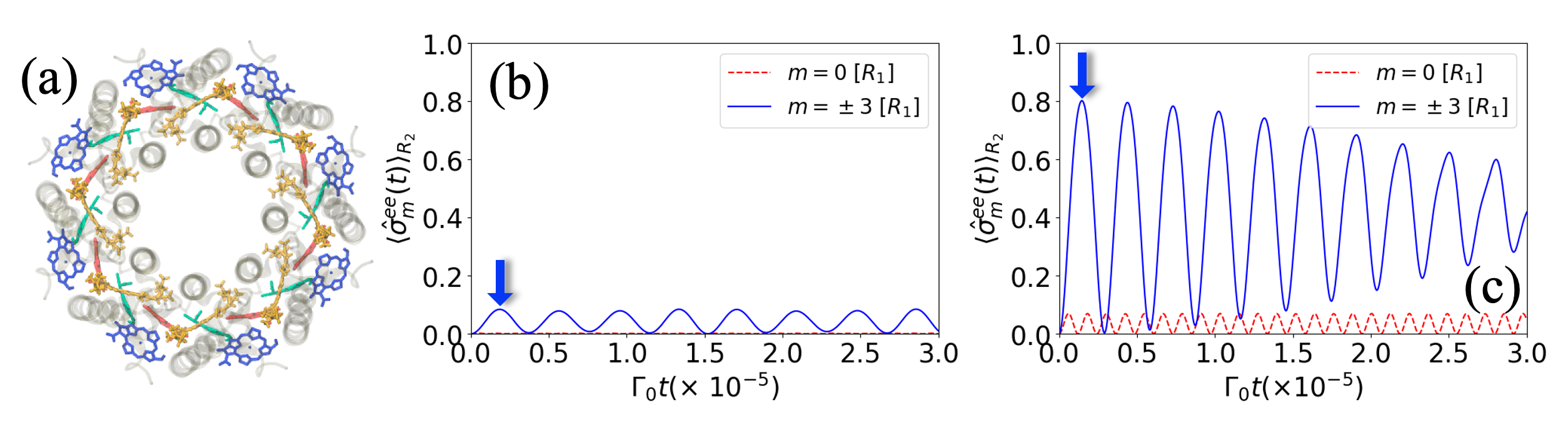}
\caption{(a) Structure of octameric LH2 of \textit{Rsp. molischianum} (top view: taken from~\cite{MALLUS:2018:CP}). BChls are color-coded as follows: blue (B800) for ${R_1}$, green (B850) for ${R_{2_a}}$, and red (B850) for ${R_{2_b}}$. The parameters utilized are listed in Table-\ref{tab4}(II), and all dipoles are assumed to be photoactive at 800 \textit{nm}. The eight-fold symmetry ($C_8$) supports seven eigenmodes. (b) For bio-size the maximum inter-layer (${R_1} \Rightarrow {R_2}$) excitation energy transfer for the darkest mode $m=\pm 3$, is only around $9\%$ (indicated by blue-arrow). (c) With a reduced vertical separation ($Z_1 = 8$ Å), the maximum transfer is significantly boosted to about 80\% (indicated by the blue arrow).}
\label{molischianum}
\end{figure}

We theoretically consider an approximate model of the LH2 complex from \textit{Rhodospirillum (Rsp.) molischianum}, which exhibits eight-fold oligomeric symmetry (see Fig.~\ref{molischianum}(a)). We approximately define geometric parameters from the reported distances in Refs.~\cite{schulten:pnas:1998, MALLUS:2018:CP}. The parameters are listed in Table-\ref{tab4}(II). For our estimations, we use the dipole-orientations of LH3 complex of \textit{Rbl. acidophilus} as reported in~\cite{Montemayor:JPCB:2018}, as a test case. Since we are interested in studying the case with all identical emitters resonant at 800 \textit{nm}, assumed to be the same as those for B800 in the LH3 complex, i.e., $|\boldsymbol{\mu}_1| = |\boldsymbol{\mu}_{2_a}| = | \boldsymbol{\mu}_{2_b}| = 6.46$D~\cite{Montemayor:JPCB:2018} for all three rings.

We calculate the approximated $C_8$ model and observe only around 9\% excitation transfer from ${R_1}$ to ${R_2}$ [Fig.~\ref{molischianum}(b)]. With a reduced vertical separation of 8 {\AA}, the modified inter-ring nearest neighbor separation will be around 9.4 {\AA}, making the unit cell nearly an equilateral triangle (intra-ring pigment separation is 9.8 {\AA}, as noted in Table-\ref{tab4}(II)). As a result, we observe a significant increase in excitation transfer around 80\% [Fig.~\ref{molischianum}(c)]. It seems that consideration of not totally accurate dipole orientation may account for not achieving the maximum excitation transfer $\sim$ 100\%, as estimated in the previous cases.


\section{Bio-mimicked stacked nanoscale ring and efficient inter-layer excitation transfer}
\label{apen-nano}
\begin{figure}
\centering
\includegraphics[width=\linewidth]{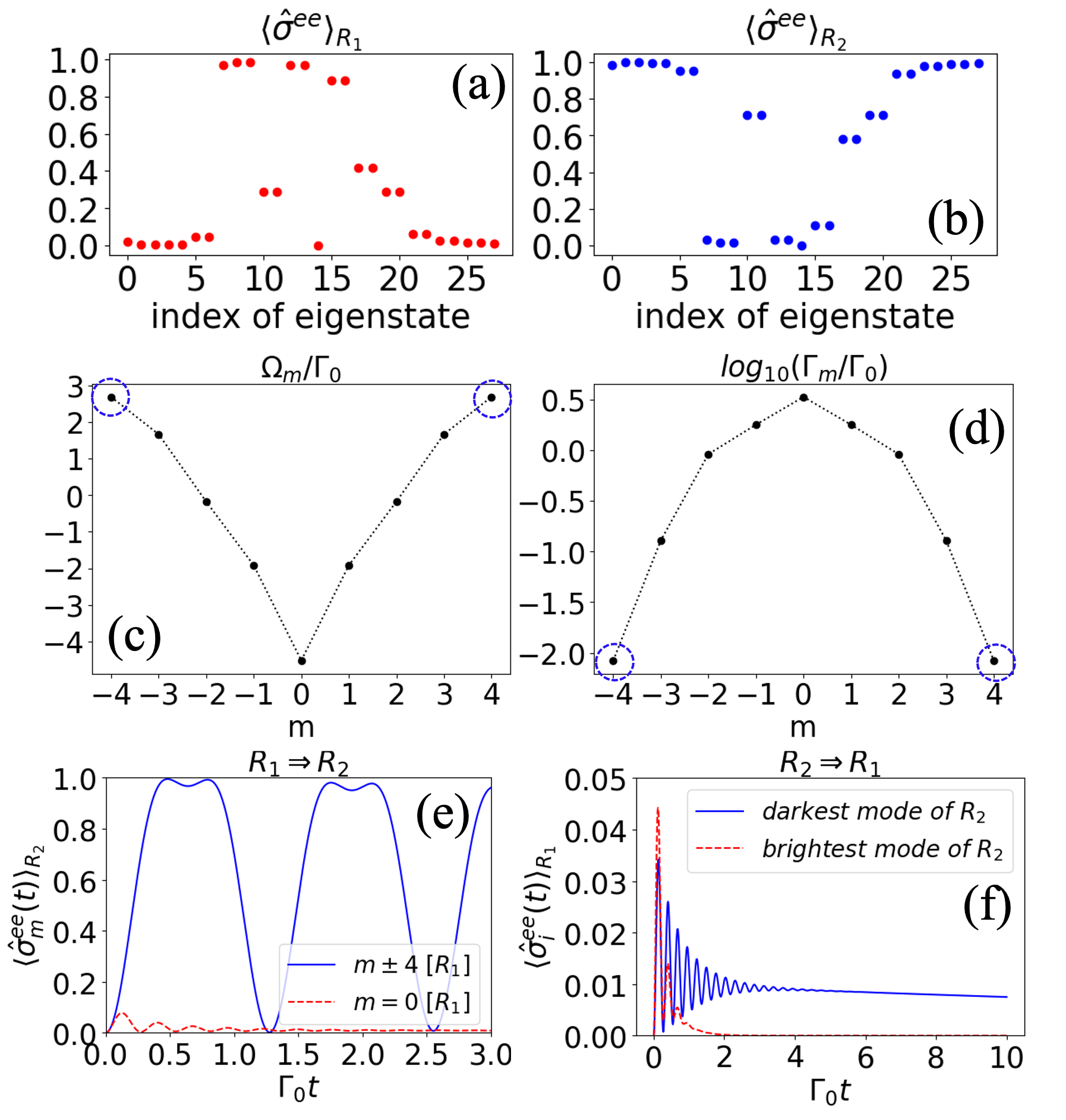}
\caption{Nanoscale stacked rings with all dipoles to be resonant to 800 \textit{nm} light and tangentially oriented ($(\theta_{1(2)},\phi_{1(2)}) = (\pi/2, \pi/2)$). Plot (a)-(b) shows population $\langle \hat n \rangle_i$ for $i^{th}$ eigenstate. Nanoscale stacked rings display strong hybridization for certain eigenstates in ring ${R_1}$ (in red) (a) and ${R_2}$ (in blue) (b). Plot (c)-(d) shows collective frequency shift ($\Omega_m/\Gamma_0$) and collective decay rate ($\Gamma_m/\Gamma_0$) for angular momentum mode $m$ of the spin-wave in ring ${R_1}$. (e) Time-evolution of the total population transferred in ${R_2}$ for eigen-modes of ring ${R_1}$ and $\Gamma_0 t \equiv 39~ns$. (f) The reverse transfer ${R_2}\Rightarrow{R_1}$ is smaller in general, here it is only around 3-4\% at maximum for the brightest and darkest modes.}
\label{fig8}
\end{figure}
Here we would briefly discuss the eigen-modes of the sparse ring ${R_1}$ of LH2-inspired nanoscale stacked concentric ring geometry of quantum emitters having transition wavelength of 800 \textit{nm} and tangentially oriented (geometric parameters are mentioned in the main text, i.e., Table-\ref{tab2}). We calculate the expectation values of the populations in ring ${R_1}$ and ${R_2}$ for each eigenstates for the Hamiltonian of Eq.(\ref{Heff}). In Fig.~\ref{fig8}(a) and (b), we show the hybridization, i.e., shared populations for some of the eigenmodes for ${R_1}$ and ${R_2}$. This hybridization is caused by the consideration of the unit-cell approaching the structure of an equilateral triangle, which eventually assists in better excitation or energy transfer as discussed in Appendix-\ref{eigen-apen}. In Fig.~\ref{fig8}(c),(d) we show the collective energy shift ($\Omega_m/\Gamma_0$) and respective collective decay rates ($\Gamma_m/\Gamma_0$) for tangentially oriented dipoles for ring ${R_1}$ only. In general, for nanorings, mode $m=0$ is the brightest, and $m=\pm 4$ is the darkest. Plot (e) displays corresponding excitation energy transfer ${R_1}\Rightarrow {R_2}$, i.e., ${\rm Max}[\langle\hat{\sigma}^{ee}_m(t)\rangle_{R_2}]$ for the above modes with scaled time $\Gamma_0t$. The darkest mode supports a maximum of 100\% excitation energy transfer (corresponds to the value 1). The time-bin here is in a few tens of nanosecond regimes. Fig.~\ref{fig8}(f) shows that the excitation transfer in the reverse direction, i.e. from ${R_2}$ to ${R_1}$ is smaller (analysis performed using the eigenstates of Hamiltonian in Eq.(\ref{Heff})). In particular, it shows only around 3-4\% of population transfer, much less than the former process. However, with some other eigenstates, we observe a bit different magnitude of excitation transfer in the reverse direction, but smaller than 100\%. Hence it seems that the inter-layer excitation transfer is more efficient from $sparse$ to $dense$ geometry, with our proposed nanoscale stacked rings in the absence of thermal decoherences.
\begin{figure}
\centering
\includegraphics[width=\linewidth]{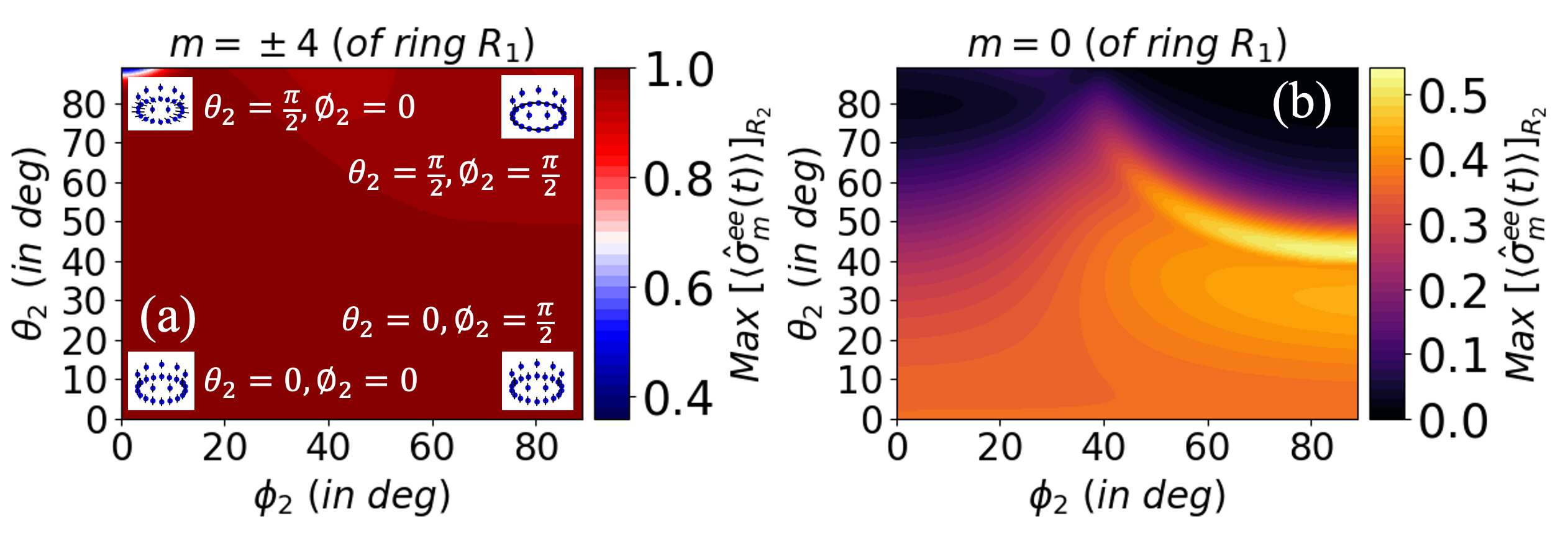}
\caption{Ring ${R_1}$ is considered to be formed by transversely orientated dipoles, i.e., $(\theta_{1},\phi_{1}) = (0^{\circ}, 0^{\circ})$. Ring sizes are the same as in Table-\ref{tab2}. For certain dipole orientations in $R_2$, the estimated maximum excitation transfer for a time-bin of $\Delta_t =$ 3$\Gamma_0t$, is 100\% for the dark mode $m= \pm 4$ (a) and $\sim$ 50 \% for the bright mode $m= 0$ (b). Example snaps for the mentioned dipole orientations in ring layers are displayed in (a).}
\label{fig9}
\end{figure}

In Fig.~\ref{fig9} we consider the dipoles of the sparse ring to be transversely orientated (magenta dotted lines in Fig.~\ref{fig1}(b)). Keeping the dipole orientation fixed in ${R_1}$, with tunable dipole orientation in the bottom ring ${R_2}$, we show the maximum of excitation energy transfer (${\rm Max}[\langle\hat{\sigma}^{ee}_m(t)\rangle_{R_1}]$) for a fixed time-bin with the darkest mode Fig.~\ref{fig9}(a) and for the brightest mode Fig.~\ref{fig9}(b). For certain choices of dipole orientations in ${R_2}$ the darkest mode enables 100\% excitation transfer and the bright mode shows a maximum of around 50\%.

\section{Excitation transfer in the stacked nanoring under the influence of static disorder and dephasing} 
\label{disorder}
\begin{figure}[b]
\centering
\includegraphics[width=0.95\linewidth]{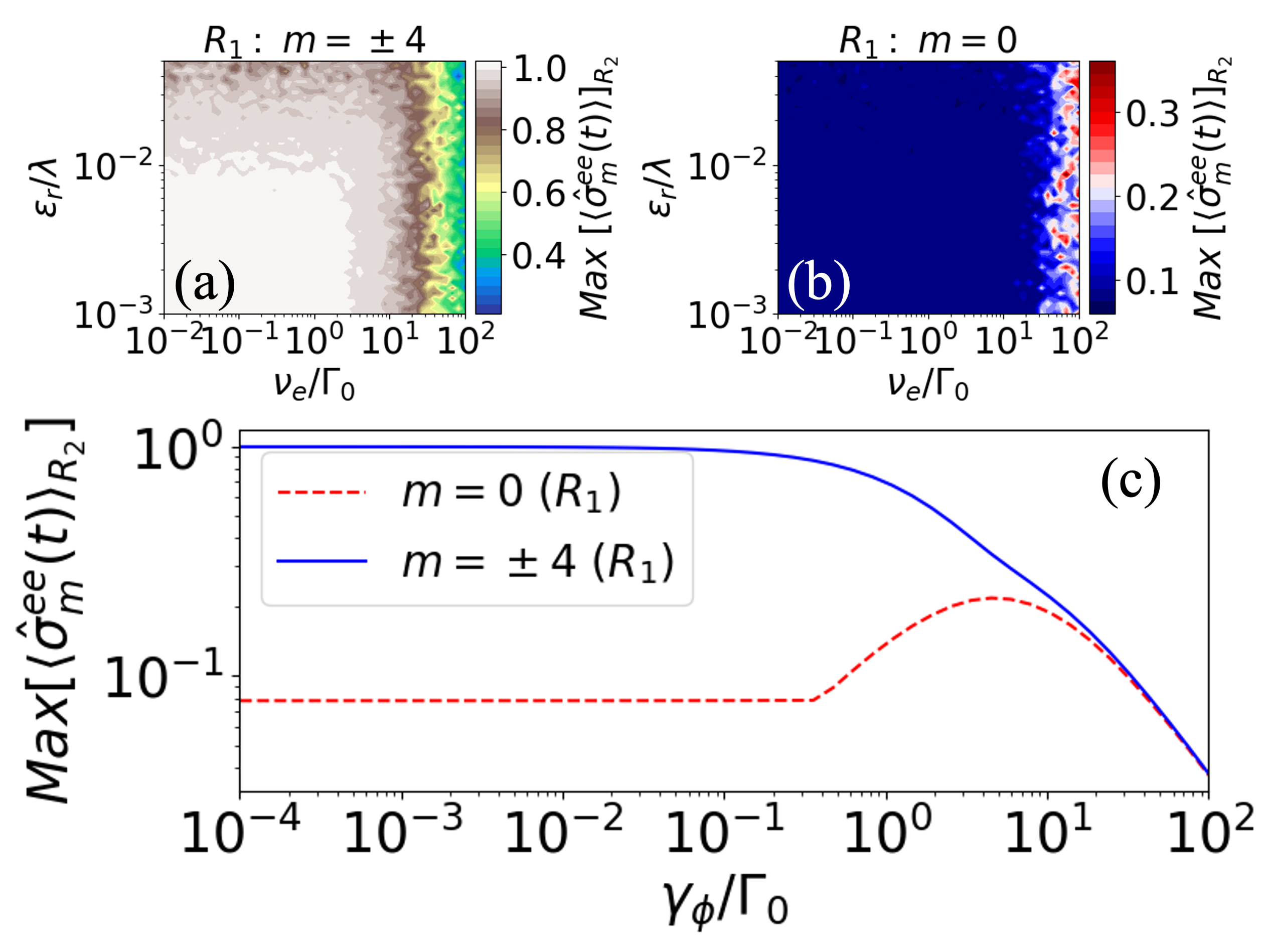}
\caption{Influence of position ($\epsilon_r$ in orders of $\lambda$) and frequency disorder ($\nu_e/\Gamma_0$) on excitation transfer ${R_1}\Rightarrow{R_2}$ in nanoscale rings; for the darkest mode $m=\pm 4$ (a) and the brightest mode $m=0$ (b) of ring ${R_1}$ (see Fig.~\ref{fig8}(c),(d)). We consider all dipoles to be in tangential orientation. In plot (c) we show the effect of dephasing ($\nu_{\phi}/\Gamma_0)$ in the excitation transfer for the above-mentioned dark and bright eigenmodes.}
\label{fig10}
\end{figure}
\textit{On-site static disorders.--} To extend our theoretical estimations at zero-temperature towards more realistic scenarios, we include the effects of static disorder of emitter positions and frequencies~\cite{kassal:jcp:2013, roos:prl:2019, harush:prr:2020,hauke:pra:2018}. For position disorder we take a normal distribution of width $\epsilon_r$ (in units of $\lambda$) in all three directions ($x,y,z$) and average over 100 realizations for each $\epsilon_r/\lambda \in \{0.001, 0.05\}$. The smallest nearest neighbor intra-inter ring distances in ${R_2} \sim 0.08 \lambda, 0.1 \lambda$ (see Table-\ref{tab3}), hence the choice of the above upper limit of $\epsilon_r/\lambda$. To account for some frequency variations $\nu_e$ (in units of $\Gamma_0$) similarly, we consider a normal distribution with a standard deviation of $\nu_e/\Gamma_0 \in \{0.01, 100\}$ and average over 100 realizations. Fig.~\ref{fig10} (a), (b) shows the variation of ${\rm Max}[\langle\hat{\sigma}^{ee}_m(t)\rangle_{R_2}]$ with position and frequency disorder in the respective axis. For the excitation transfer of the darkest modes $m=\pm 4$ we see a small continuous decrease of efficiency with disorder (a). However, disorder mixes the bright mode and the subradiant modes and thus reduces free space emission for modes with $m \approx  0$. Hence this leads to an increase in the efficiency of coupling and decay time scales roughly match. A similar behavior was observed for ring structures including vibrational degrees of freedom~\cite{holzinger2022cooperative}. For much stronger disorder, both, free space loss and transfer are hampered and thus this reduces the overall transfer efficiency again.

{\textit{Influence of dephasing.--}} We also consider the effect of local dephasing of each emitter due to its local environment (for instance, Dibenzoterrylene molecules in a substrate~\cite{Hood:np:2024}) as a rough model of phonons or motion, which can be incorporated via the following modification in the master equation~\cite{Plenio_2008, Rebentrost_2009} 
\bea
\dot{\rho} = -i[H_{\rm DD}, \rho] + \mathcal{L} [\rho] + 2\gamma_{\phi} \sum_i \hat{\sigma}^+_i\hat{\sigma}^-_i \rho \hat{\sigma}^+_i\hat{\sigma}^-_i~,
\eea
where $\gamma_\phi$ is the local dephasing rate of the emitters. We solve the above equation to determine the maximum excitation energy transfer via Eq.(\ref{exc-tran}) in the bottom ring with the modified wave function over a time-bin $\Delta_t$. Interestingly in plot Fig.~\ref{fig10}(c), we witness that with some optimum value of $\gamma_{\phi}/\Gamma_0$ the excitation transfer increases for the bright mode $m=0$, then after a certain point it decreases and merges with the dark modes $m=\pm 4$. As mentioned above, dephasing also assists in increasing the excitation transfer~\cite{Plenio_2008, Alan:JCP:2008, Chin:Njp:2010, plenio:NJP:2014, roos:prl:2019,harush:prr:2020, peter:prr:2024}. It does so by starting from the bright symmetric mode $m = 0$, where it suppresses superradiance as a competing effect, and by transferring some population to the dark state manifold, where transport is enhanced. For this to be effective, dephasing needs to be strong enough to prevent superradiant decay but weak enough to still facilitate coherent dark transport.


\bibliography{draftbib}

\end{document}